\documentclass[longauth]{aa} % for the long lists of affiliations 
\pdfoutput=1

\usepackage{natbib}
\bibpunct{(}{)}{;}{a}{}{,} % to follow the A&A style

\usepackage{graphicx}
\usepackage{lscape}
\usepackage{threeparttable}
\usepackage{txfonts}
\usepackage{subcaption}
\begin{document}

   \title{The Gaia-ESO Survey: Age-chemical-clock relations spatially resolved in the Galactic disc \thanks{Based on observations collected with the FLAMES instrument at
VLT/UT2 telescope (Paranal Observatory, ESO, Chile), for the Gaia-
ESO Large Public Spectroscopic Survey (188.B-3002, 193.B-0936, 197.B-1074).} }
\titlerunning{Chemical clocks in 2D} 
\authorrunning{Viscasillas Vázquez et al.}

   \author{C. Viscasillas Vázquez\inst{\ref{vilnius},\ref{oaa}}, 
   L. Magrini\inst{\ref{oaa}}, G. Casali\inst{\ref{difa},\ref{oabo}},
   G. Tautvai\v{s}ien\.{e}\inst{\ref{vilnius}}, 
   L. Spina \inst{\ref{oapd}}, 
   M. Van der Swaelmen\inst{\ref{oaa}},   
   S. Randich\inst{\ref{oaa}},  
   T. Bensby\inst{\ref{lund}}, 
   A. Bragaglia\inst{\ref{oabo}},  
   E. Friel\inst{\ref{indiana}}, 
   S. Feltzing\inst{\ref{lund}}, 
   G.G. Sacco\inst{\ref{oaa}},
   A. Turchi\inst{\ref{oaa}}, 
   F. Jim\'enez-Esteban\inst{\ref{cab}}, 
   V. D'Orazi\inst{\ref{oapd}}, 
   E. Delgado-Mena\inst{\ref{porto}},
   %here people who want to be author, but not giving major contribution
    Š. Mikolaitis\inst{\ref{vilnius}}, 
    A. Drazdauskas\inst{\ref{vilnius}}, 
    R. Minkevičiūtė\inst{\ref{vilnius}}, 
    E. Stonkutė\inst{\ref{vilnius}}, 
    V. Bagdonas\inst{\ref{vilnius}}, 
    D. Montes\inst{\ref{ucm}},  
    G. Guiglion\inst{\ref{postdam}}, 
    M. Baratella\inst{\ref{postdam}},  
    H. M. Tabernero\inst{\ref{cab2}}, 
   %from here only GES from the builder list (no contribution to the specific paper)
   G. Gilmore\inst{\ref{cambridge}},  
   E. Alfaro\inst{\ref{iaa}},  
   P. Francois\inst{\ref{gepi}}, 
   A. Korn\inst{\ref{upsala}}, 
   R. Smiljanic\inst{\ref{poland}}, 
   M. Bergemann\inst{\ref{mpi}}, 
   E. Franciosini\inst{\ref{oaa}}, 
   A. Gonneau\inst{\ref{cambridge}},   
   A. Hourihane\inst{\ref{cambridge}}, 
   C.~C. Worley\inst{\ref{cambridge}},      
   S. Zaggia\inst{\ref{oapd}} 
          }

\institute{
Institute of Theoretical Physics and Astronomy, Vilnius University, Sauletekio av. 3, 10257 Vilnius, Lithuania \email{carlos.viscasillas@ff.vu.lt} \label{vilnius} 
\and
INAF - Osservatorio Astrofisico di Arcetri, Largo E. Fermi 5, 50125, Firenze, Italy \email{laura.magrini@inaf.it} \label{oaa} 
\and
Dipartimento di Fisica e Astronomia, Università degli Studi di Bologna, Via Gobetti 93/2, I-40129 Bologna, Italy\label{difa}
\and
INAF - Osservatorio di Astrofisica e Scienza dello Spazio di Bologna, via Gobetti 93/3, 40129, Bologna, Italy\label{oabo} 
\and
INAF - Padova Observatory, Vicolo dell'Osservatorio 5, 35122 Padova, Italy\label{oapd} 
\and
Lund Observatory - Department of Astronomy and Theoretical Physics, Box 43, SE-22100 Lund, Sweden\label{lund}
\and
Department of Astronomy - Indiana University, Swain West 318
727 East Third Street, Bloomington, IN 47405, USA\label{indiana}
\and
Departamento de Astrof\'{\i}sica, Centro de Astrobiolog\'{\i}a (CSIC-INTA), ESAC Campus, Camino Bajo del Castillo s/n, E-28692 Villanueva de la Ca\~nada, Madrid, Spain\label{cab}
\and
Instituto de Astrofísica e Ciências do Espaço, Universidade do Porto, CAUP, Rua das Estrelas, 4150-762 Porto, Portugal\label{porto}
\and
Departamento de Física de la Tierra y Astrofísíca and IPARCOS UCM, Universidad Complutense de Madrid, E-28040, Madrid, Spain\label{ucm}
\and
Leibniz-Institut für Astrophysik Potsdam, An der Sternwarte 16, 14482 Potsdam, Germany\label{postdam}
\and
 Centro de Astrobiología (CSIC-INTA), Carretera de Ajalvir km 4, Torrejón de Ardoz, 28850, Madrid, Spain\label{cab2}
\and
  Institute of Astronomy, University of Cambridge, Madingley Road, Cambridge CB3 0HA, United Kingdom\label{cambridge}
  \and
Instituto de Astrofísica de Andalucía, CSIC, Glorieta de la Astronomía, E-18080, Granada, Spain\label{iaa}
   \and
  GEPI, Observatoire de Paris, CNRS, Universit\'e Paris Diderot, 5 Place Jules Janssen, 92190 Meudon, France\label{gepi}
  \and
Observational Astrophysics, Department of Physics and Astronomy, Uppsala University, Box 516, 75120 Uppsala, Sweden\label{upsala}
\and
Nicolaus Copernicus Astronomical Center, Polish Academy of Sciences, ul. Bartycka 18, 00-716, Warsaw, Poland\label{poland}
\and
Max-Planck Institut für Astronomie, Königstuhl 17, 69117 Heidelberg, Germany\label{mpi}
}

   \date{Received 17 December 2021/ Accepted 25 January 2022 }

  \abstract
   {The last decade has seen a revolution in our knowledge of the Galaxy thanks to the {\em Gaia} and asteroseismic space missions and  the ground-based spectroscopic surveys.
   }
   { To complete this picture, it is necessary to map the ages of its stellar populations. During recent years, the dependence on time of abundance ratios involving slow (s) neutron-capture and $\alpha$ elements (called chemical-clocks) has been used to provide estimates of stellar ages, usually in a limited volume close to the Sun. We aim to analyse the relations of chemical clocks in the Galactic disc extending the range to R$_{\rm GC}\sim$6-20~kpc. }
   {Using the sixth internal data release of the {\em Gaia}-ESO survey, we calibrated  several relations between stellar ages and abundance ratios  [s/$\alpha$] using a sample of open clusters, the largest one so far used with this aim (62 clusters). Thanks to their wide galactocentric coverage, we investigated the radial variations of the shape of these relations, confirming their non-universality.  }
   {The multi-variate relations allowed us to infer stellar ages for field stars. We estimated our accuracy (ranging from 0.0 to -0.9 Gyr) and precision (from 0.4 to 2.3 Gyr) in recovering the global ages of open clusters, and the ages of their individual members. We applied the relations with the highest correlation coefficients to the field star population, finding an older population at lower metallicity and higher [$\alpha$/Fe] in the thin
disc, and a younger one at higher [Fe/H] and low [$\alpha$/Fe], as expected. 
   }
   {We confirm that there is no single age-chemical clock relationship valid for the whole disc, but that there is a dependence on the galactocentric position, which is related to the radial variation of the star formation history combined with the non-monotonic dependence on metallicity of the yields of the $s$-process elements from low- and intermediate-mass stars. Finally, the abundance ratios [Ba/$\alpha$] are more sensitive to age than those with [Y/$\alpha$] for young disc stars, and their slopes vary less with galactocentric distance.
   We remind the reader that the application of such relationships to field stars is only of statistical value. }  


   \maketitle
%
%-------------------------------------------------------------------

\section{Introduction}

    In the multi-dimensional space traced by {\it Gaia} \citep{gaia16, gaia18, gaia21} and the large, high-resolution spectroscopic surveys, such as {\it Gaia}-ESO \citep{Gilmore_2012, Randich_2013}, APOGEE \citep{majewski17}, and GALAH \citep{desilva15}, the missing variable is time, which can be traced by the ages of stars. 
    However, stellar ages, as the other properties of stars,  cannot be directly measured. They need a direct comparison with the outputs of stellar evolution models, through the so-called isochrone fitting, in which observed (colours and magnitudes) or derived (temperature and surface gravity) quantities are compared with theoretical values. Moreover, having both good models of stellar evolution and good observational measurements is not always sufficient to infer reliable age estimates, since  isochrone fitting gives unsatisfactory results for field stars, particularly on the main sequence and the giant branch in  of the Hertzsprung-Russel (HR) diagram, where the isochrones are particularly crowded. Therefore, in recent years, alternative or complementary methods have been put in place to provide further estimates of stellar ages \citep[see, e.g.][]{soderblom14, howes19}. 
    
    The first study, to our knowledge, to notice the net increase in the abundance of slow (s) neutron capture elements in young stellar populations is \citet{dorazi09}, in which the abundance of barium in young star clusters was seen to be higher than in the older ones.  
    \citet{maiorca11,maiorca12} added a few more elements with important $s$-process contributions (yttrium, zirconium, lanthanum, and cerium), confirming the increasing trend towards younger ages. Subsequently, a number of works have attempted to both clarify the origin of this increase \citep[see, e.g.][]{Bisterzo14, Mishenina15, trippella16, magrini18, spina18, busso21} and to use their abundances to estimate the ages of stars, often using neutron capture $s$-process elements in combination with other elements with opposite behaviours, such as $\alpha$ elements -- that we indicate as chemical clocks -- and thus maximising the dependence of the relationship with age \citep[see, e.g.][]{tucci16, Nissen16, feltzing17, Fuhrmann17, slumstrup17, titarenko19}. 
    Once the existence of a relationship between age and chemical clocks was established \citep[see, e.g.][]{spina16, delgado19, jofre20}, the next steps were the following:  i) to clarify the applicability of these relationships with luminosity class (dwarf or giant) \citep[see, e.g.][]{tuccimaia16, slumstrup17, casamiquela21}, metallicity \citep[see, e.g][]{feltzing17, delgado19, casali20}, and population type (thin disc, thick disc, halo) \citep[see, e.g.][]{titarenko19, nissen20, tautvasiene21}, or even in dwarf galaxies \citep{skuladottir19}; ii) to calibrate them with a sample of stars with reliable age determination, which are usually open star clusters (OCs), solar twins, or targets with asteroseismic observations. 
    Finally, it is essential to understand whether these relationships are valid throughout the Galactic disc, or whether they are necessarily position-dependent.
    For the first time, \citet{casali20} applied the relations derived from a large sample of solar-like stars located in the solar neighbourhood and noted that they fail to reproduce the ages of star clusters in the inner disc. They concluded that the relationship between age and chemical clocks is not universal and that it varies with galactocentric position. 
    Later, \citet{magrini21a} suggested that the differences in the relationships between age and chemical clocks in different parts of the Galactic disc are due to the strong dependence on the metallicity of the yields of low-mass stars, which produce $s$-process elements during the final stages of their evolution.
    \citet{casamiquela21} used red clump stars in open clusters to investigate the age dependence of several abundance ratios, including those that contain $s$-process and $\alpha$ elements.  They found that the relationship between  [Y/Mg] and ages outlined by open clusters is 
    similar to the one found using solar twins in the solar neighbourhood. They also found that the abundance ratios involving Ba are those with the highest correlation with age. However, they also note that as one moves away from the solar neighbourhood, the dispersion increases and is in agreement with the findings of \citet{casali20}, which attributed this to the spatial variation of the star formation history along the galactocentric radius. 
    
    In the present paper, we discuss the radial variation of the relations between ages and chemical clocks using the largest sample of open clusters so far used for this purpose: 62 OCs observed by the {\em Gaia}-ESO survey sixth internal release, {\sc idr6}, which includes science clusters, calibration clusters, and archive clusters, that is those which were not observed by {\em Gaia}-ESO but were present in the ESO archive and then homogeneously re-analysed by the {\em Gaia}-ESO consortium. 
    The paper is structured as follows. In Section~\ref{sec_data}, we describe the {\em Gaia}-ESO database and the solar-scale normalisation, while in Section~\ref{sec_samples} we depict our samples of member stars in open clusters and of field stars. 
    In Section~\ref{sec_multivariate}, we describe the relationships between [El/Fe] and age and between the chemical clocks and ages, using open star clusters as calibrators. We also estimate the role of migration in determining such relations. We apply our relations to cluster stars (Section~\ref{sec_ages_clusters}) and to field stars (Section~\ref{ages_fiels}). 
    In Section~\ref{sec_discussion}, we discuss our results and provide our conclusions.

\section{Data reduction  and analysis }
\label{sec_data}

\subsection{Tha Gaia-ESO {\sc idr6}}
We used data from {\sc idr6} of the Gaia-ESO Survey obtained from the spectral analysis from the UVES spectra (resolving power R$=$47,000 and spectral range 480.0$-$680.0~nm). The spectra were reduced and analysed by the {\em Gaia}-ESO consortium, and the data analysis is organised in different working groups (WGs). The spectra are analysed with several pipelines, internally combined by each WG, and then the results from the different WGs are homogenised using a database of calibrators, such as benchmark stars and open or globular clusters selected following the calibration strategy by \citet{pancino17} and adopted for the homogenisation by WG15 (Hourihane et al. in preparation). 
The data reduction, radial and rotational velocity determinations are undertaken in INAF-Arcetri \citep{sacco14} using the FLAMES-UVES ESO public pipeline. The analysis process for the UVES spectra of F, G, and K stars in the field of the Milky Way (MW), in open clusters, and in calibration targets, including globular clusters,  was described in \citet{smi14}. 
 
The recommended parameters and abundances are distributed in the {\sc idr6} catalogue, which includes those used in the present work: atmospheric stellar parameters, such as effective temperature,  $T_\mathrm{eff}$, surface gravity,  $\log g$, and metallicity, [Fe/H], and abundances of 32 elements, several of them in their neutral and ionised forms. 
In the present paper, we discuss the abundances of five $s$-processes,  four $\alpha$ elements, and an odd-Z element.  All the abundances (by number) are expressed in their usual logarithmic form: A(El)=12+$\log$(n(El)/n(H)), while the abundances normalised to the solar scale are indicated with [El/H]= $\log$(n(El)/n(H))-$\log$(n(El)/n(H))$_{\odot}$).

\subsection{Solar abundance scale}
In Table~\ref{solarabundance}, we show, for the elements used throughout this work, both the solar abundances derived for the Sun in {\em Gaia}-ESO {\sc idr6} and those from \citet{Grevesse_2007}, which are used in  {\it Gaia}-ESO to build up both the model atmospheres and the synthetic spectra used by the consortium. We also include the average abundances of M67, which has a chemical composition very similar to the solar one \citep[see, e.g.][]{onehag11, liu16}, and the abundances of their giant and dwarf member stars observed in {\em Gaia}-ESO {\sc idr6}. We separate dwarf and giant stars in M~67 on the basis of their surface gravity: we define a giant star if $\log g < 3.5$, and a dwarf star if $\log g \geq 3.5$.
In the first part of Table~\ref{solarabundance}, we present the adopted $\alpha$ and odd-Z elements, namely Mg, Al, Si, Ca, and Ti. 
In the second part, we show the slow neutron-capture elements (Y, Zr, Ba, La, Ce). 
The agreement between the solar and the average M67 {\em Gaia}-ESO abundances, and those of \citet{Grevesse_2007}, is very good, within 1-$\sigma$. 
In the last two columns, we show the abundances of the giant and dwarf members of M67 separately, to estimate the impact of a single set of normalising values to both giant and dwarf stars. \citet{bertellimotta18} investigated variations in the surface chemical composition of member stars of M67 as a possible consequence of atomic diffusion, which  takes place during the main-sequence (MS) phase \citep[see, e.g.][]{bertellimotta17, souto19}.
They found that  the abundances of MS stars differ with respect to those of the giant stars, consistently with the predictions of stellar evolutionary models. 
In our sample, these differences and other possible effects due to spectral analysis are also noted on the average values of M67 dwarfs and giants, in particular  for Al, and for Si among the odd-Z and $\alpha$-elements and Y among the neutron capture.

Although these variations are not very large, they can affect our comparison between giant stars (essentially cluster members) and dwarf stars (most of them composing the field star sample), modifying their abundance scale. We thus normalised the abundances of dwarf and giant stars using two different values, corresponding to average values of M67 for dwarf and giant stars (fifth and sixth columns of Table~\ref{solarabundance}).

\begin{table*}
\caption{{\sc iDR6} solar and M67 abundances for neutron-capture elements and  $\alpha$-elements.}              % title of Table
\label{solarabundance}      % is used to refer this table in the text
\centering                                      % used for centering table
\begin{tabular}{l c c c c c}          % centered columns (4 columns)
\hline\hline                        % inserts double horizontal lines
A(El) & Sun ({\sc iDR6}) & Sun & M67 ({\sc iDR6}) & M67 ({\sc iDR6}) & M67 ({\sc iDR6}) \\ 
& &  \citet{Grevesse_2007} & & (giants) & (dwarfs) \\ % table heading
\hline                                   % inserts single horizontal line
   
   Mg I & 7.51 $\pm 0.02$& 7.53 $\pm 0.09$ & 7.50 $\pm 0.04$  & 7.53 $\pm$ 0.04  & 7.49 $\pm$ 0.04\\
   Al I & 6.40 $\pm 0.02$& 6.37 $\pm 0.06$ & 6.42 $\pm 0.03$  & 6.49 $\pm$ 0.03 & 6.39 $\pm$ 0.03\\
   Si I & 7.44 $\pm 0.02$& 7.51 $\pm 0.04$ & 7.44 $\pm 0.03$  & 7.50 $\pm$ 0.02 & 7.41 $\pm$ 0.02\\
%   Si II & 7.52 $\pm 0.02$& 7.51 $\pm 0.04$ & 7.46 $\pm 0.05$  & 7.59 $\pm$ 0.05 & 7.46 $\pm$ 0.05\\
   Ca I & 6.32 $\pm 0.02$& 6.31 $\pm 0.04$ & 6.26 $\pm 0.02$  & 6.25 $\pm$ 0.02 & 6.29 $\pm$ 0.03\\
%   Ca II &  6.30 $\pm 0.02 $ & 6.31 $\pm 0.04$ & 6.26 $\pm 0.02$  & 6.36 $\pm 0.02 $ & 6.23 $\pm $ 0.02 \\
   Ti I & 4.89 $\pm 0.02$& 4.90 $\pm 0.06$ & 4.90 $\pm 0.03$  & 4.87 $\pm$ 0.03 & 4.91 $\pm$ 0.03\\
%   Ti II & 4.95 $\pm 0.02$& 4.90 $\pm 0.06$ & 4.94 $\pm 0.03$  & 4.94 $\pm$ 0.02 & 4.94 $\pm$ 0.04\\
   \hline 
   Y II & 2.19 $\pm 0.04$ & 2.21 $\pm 0.02$ & 2.17$\pm0.04$ & 2.09 $\pm$ 0.02 & 2.21 $\pm$ 0.05\\
   Zr I & 2.62 $\pm 0.13$ & 2.58 $\pm 0.02$ & 2.54$\pm0.02$ & 2.51 $\pm$ 0.03 & 2.61 $\pm$ 0.02\\    
%   Zr II & - & 2.58 $\pm 0.02$ & 2.53$\pm 0.04$ & 2.57 $\pm $ 0.04 & 2.51 $\pm$ 0.04\\    
%   Mo I & 2.01 $\pm 0.06$ & 1.92 $\pm 0.05$ & 1.92$\pm 0.03$ & 1.92 $\pm$ 0.03 & - \\    
   Ba II & 2.20$\pm0.04$ & 2.17 $\pm 0.07$ & 2.17$\pm 0.06$ & 2.12 $\pm$ 0.07 & 2.19 $\pm$ 0.06\\    
   La II & 1.13$\pm 0.02$ & 1.13 $\pm 0.05$ & 1.17$\pm 0.07$ & 1.17 $\pm$ 0.02 & 1.17 $\pm$ 0.09\\    
   Ce II & 1.70$\pm 0.02$ & 1.70 $\pm 0.10$ & 1.66$\pm 0.07$ & 1.66 $\pm$ 0.03 & 1.67 $\pm$ 0.09\\    
%   Pr II & 0.57$\pm 0.02$ & 0.58 $\pm 0.10$ & 0.55$\pm 0.04$ & 0.54 $\pm$ 0.03 & 0.55 $\pm$ 0.06\\    
%   Nd II & 1.49$\pm0.02$ & 1.45 $\pm 0.05$ & 1.48$\pm0.06$ & 1.41 $\pm$ 0.03 & 1.49 $\pm$ 0.07\\  
   %Sm II & - & 1.00 $\pm0.03$ & - \\ 
%   Eu II &0.52$\pm0.02$  & 0.52 $\pm 0.06$ & 0.52$\pm 0.05$ & 0.53 $\pm$ 0.04 & 0.53 $\pm$ 0.05\\  
    
\hline                                             %inserts single line
\end{tabular}
\end{table*}
 
%7

%-----------------------------------------------------------------

\section{Samples}
\label{sec_samples}
The sample of WG~11 stars in {\sc idr6} is made up of 6877 stars (column 'WG' in the {\em Gaia}-ESO nomenclature), of which 3613 are Galactic field stars, 2794 are classified as star cluster candidates, spread over 96 (both globular and open) clusters, and 470 are standard field stars, including benchmark stars and asteroseismic fields. 
We divided our sample into two sets, the first containing only members of clusters older than 100 Myr, and the second containing the field stars. We used the first sample to calibrate the relations between age, galactocentric distance, R$_{\rm GC}$,  and abundance ratios, while the second sample was used to cross-check the applicability of our method.

\subsection{Sample of member stars of open clusters}\label{clusters}

\begin{figure}
  \resizebox{\hsize}{!}{\includegraphics{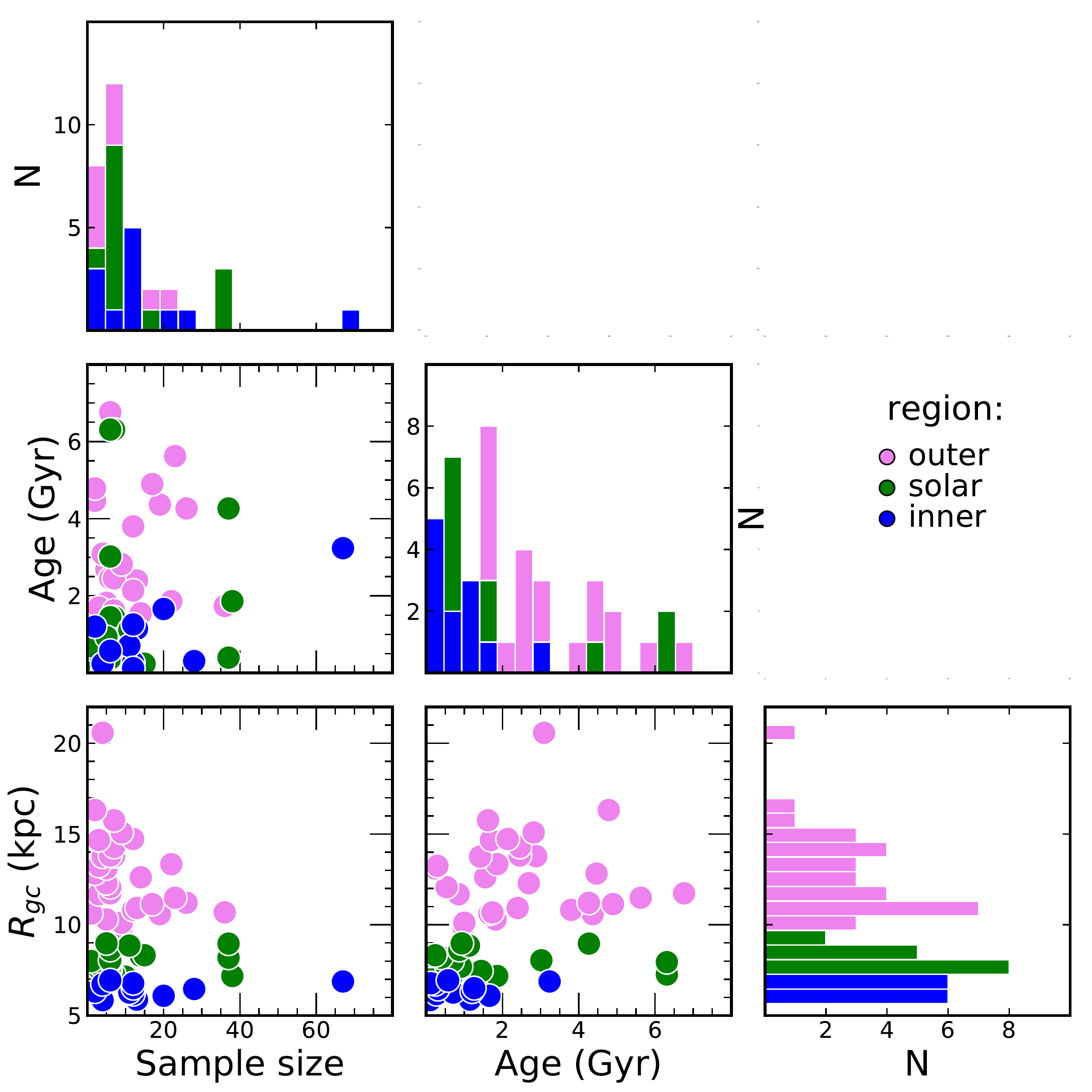}}
  \caption{\label{age_rgc_distribution} Distributions of ages, galactocentric distances, and sizes (number of observed members in {\sc dr6}) for our sample of clusters colour-coded by R$_{\rm GC}$ (inner disc R$_{\rm GC} <$7~kpc in blue; solar region 7~kpc$<$R$_{\rm GC}<$9~kpc in green; outer disc R$_{\rm GC} >$9~kpc in pink).
  }
\end{figure}

Open clusters  are among the best calibrators between the properties of stars and their ages. The member stars of the same cluster show homogeneity in age and chemical composition for most elements \citep[exceptions are those processed within the nuclear region of the star and brought to the surface by convection during the evolved phases of giants such as C, N, and Li; see, e.g. ][]{randich21} and within the effect related to a secondary process such as atomic diffusion. 

\begin{figure}
  \resizebox{\hsize}{!}{\includegraphics{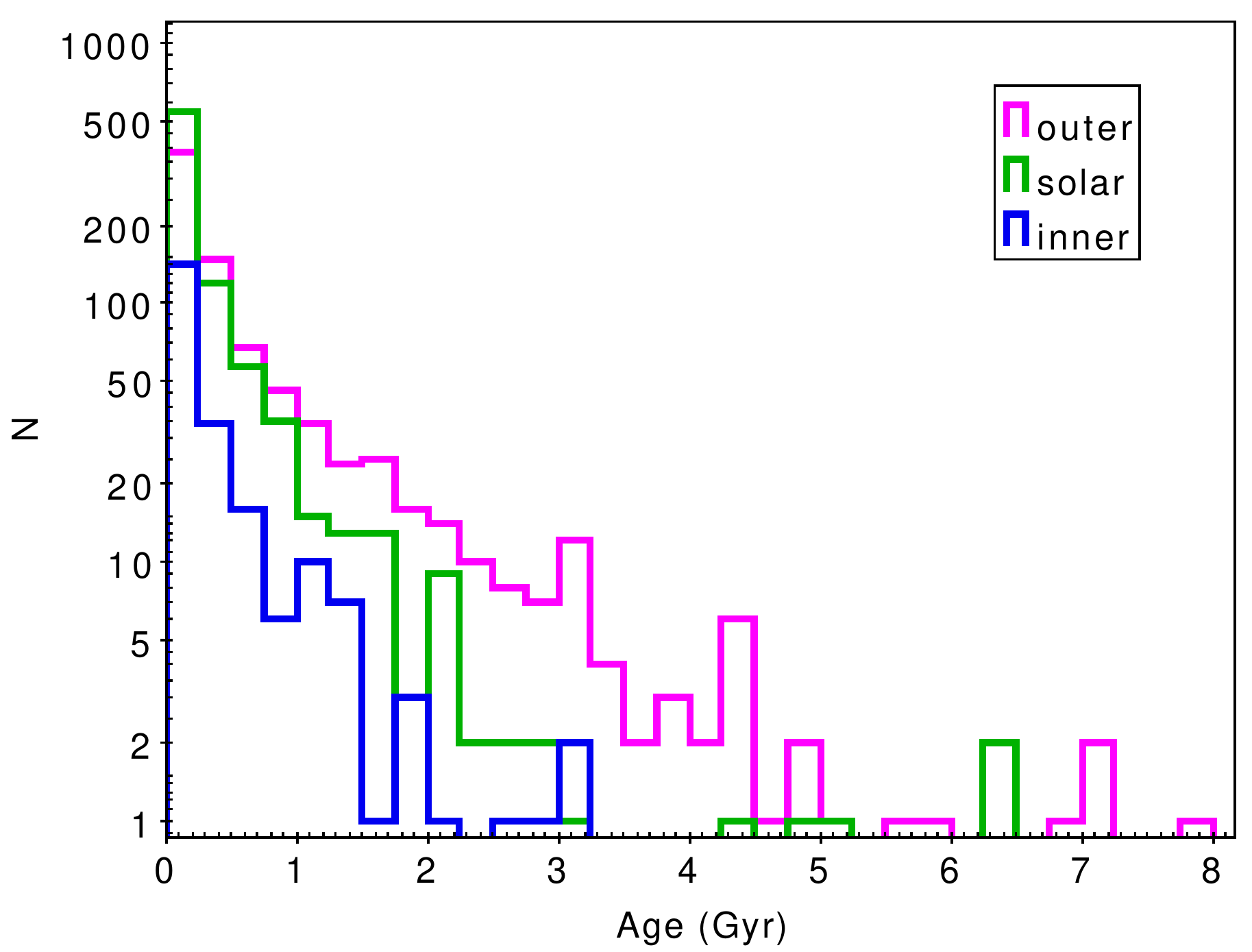}}
  \caption{\label{age_rgc_distribution_cg20} Distributions of ages in the three radial bins for the sample of 2017 open star clusters of \citet{CG20}. The histograms are colour-coded by R$_{\rm GC}$ (inner disc R$_{\rm GC} <$7~kpc in blue; solar region 7~kpc$<$R$_{\rm GC}<$9~kpc in green; outer disc R$_{\rm GC} >$9~kpc in pink).
  }
\end{figure}

In the present work, we considered 62 OCs with age$\geq$100~Myr, which contain a total number of 788 member stars.  
Among the cluster members, we only considered those that have at least one abundance value of an $s$ process and of one of the considered  $\alpha$- or odd-Z elements, which reduced our sample to 716 stars. The analysis of younger clusters should be specific, as detailed in \citet{Baratella_2020, baratella21}, and the determination of their abundances are affected by activity, rotation, and problems in the derivation of the microturbulent velocity, $\xi$. For this reason, they were not included in this work. 
For our sample clusters, we adopted ages and galactocentric distances from the homogeneous analysis  with {\em Gaia} {\sc dr2} data of \citet{CG20}. 
In Figure~\ref{age_rgc_distribution}, we show the distributions of their number of observed member stars (sample size), age, and R$_{\rm GC}$. Since we only used the UVES data, for most clusters we have about ten members, but in some cases there are more. The ages range from 100 Myr to about 7 Gyr, while the R$_{\rm GC}$ from about 6~kpc to 20~kpc. 
An interesting aspect of the open cluster sample is the absence of clusters older than about 3 Gyr in the inner disc. 
A comparison with the age distribution in the same three radial intervals used in our work of the whole cluster sample in \citet{CG20} clearly shows the same effect: an almost total absence of old clusters in the inner disc, as shown in Figure~\ref{age_rgc_distribution_cg20}. 
The limitation is likely intrinsic, due to the higher efficiency of destructive processes in higher density areas, as recently observed in M51. \citet{messa18} found that the age distribution of clusters is dependent on the region considered and is consistent with rapid disruption only in dense regions, while little disruption is observed at large galactocentric distances and in the inter-arm region.

This lack of inner old clusters has significant consequences on our calibration work, because it will not allow us to date the oldest stars in the inner disc unless we accept extrapolations of the relationships.

 The membership and the corresponding probability are computed in different ways depending on how many stars are observed in each cluster. 
For 41 clusters (676 stars), we used the membership analysis described in \citet{jackson22}, and for the remaining clusters we used the analysis of \citet{magrini21b}. 
The former used a maximum likelihood technique to determine membership probabilities for each star based on their 3D kinematics, combining the parameters of {\em Gaia}-ESO with {\em Gaia} {\sc edr3}, 2MASS, and VISTA. 
Using the membership of \citet{jackson22}, we only included members with probability $>$ 0.9 (MEM3D).
For the other clusters, following \citet{magrini21b}, the selection of members was done in two different ways according to the number of observed candidate stars. When the cluster had more than 20 potential member stars, the selection was done by performing a simultaneous fit of the {\em Gaia}-ESO radial velocities, RVs,  and the parallaxes and proper motions from Gaia {\sc edr3} \citep{gaia21}. When the cluster had fewer than 20 potential member stars, the authors derived the peak and standard deviation of the RV distribution and selected stars within 2$\sigma$ of the peak, then computing the average parallax, proper motion, and their standard deviations, and excluding those differing by more than 2$\sigma$ from the average values.

Since the median signal-to-noise ratio (S/N) of the cluster member stars ($<$S/N$\sim$100) is higher than that of field stars, and since we did not use individual abundances of member stars for clusters, but we adopted average values from all members, we did not apply any selection criteria based on the S/N or errors in parameters. We only discarded some stars with high errors in abundance of neutron capture elements (error in abundance $\geq$ 0.1 dex).
In addition, for each cluster, we used the interquartile range rule to detect potential outliers that fall outside of the overall abundance pattern. This range is defined by (Q1 - 1.5$\times$IQR, Q3 + 1.5$\times$IQR), where IQR is the difference between the 75th and 25th percentiles of the data, being Q1 the lower quartile, Q2 the median, and Q3 the upper quartile (see Figure~\ref{fig:outliers} in which the outlier stars stand out from the main distribution for each cluster).

We examined the outliers for each cluster individually and in the context of the entire data set, finding some stars with anomalous abundances, in several cases in extreme abundance of $s$-process elements (e.g. 07465200-0441557 in Berkeley 39, with differences above 4 $\sigma$ for [Y/H] and [Ba/H], 06025078+1030280 in NGC2141, and 06071407+2406547 in NGC2158, where those IDs are the CNAMEs given in the {\em Gaia}-ESO survey). The exclusion of out of range stars (26 stars that are outliers in more than one $s$-process element) in addition to the reduction of the scatter improves the correlations. 
These stars will be analysed in detail in a future work, in which the overabundance of s-elements will be correlated with stellar properties, including binarity and rotation.

In Table 1 of the appendix, we provide the global metallicity of each cluster from \citet{Randich_2022}, together with R$_{\rm GC}$ and age \citep{CG20}, and the abundance ratios used along the paper. We provide both [El/H] and [El/Fe] because the transition between the two ratios is not straightforward, as they are calculated star by star and then averaged, while the overall metallicity [Fe/H] is generally calculated with a larger number of members.

\subsection{Sample of field stars}\label{sec:sample:mw}
Our set of field stars is composed of two sub-samples: {\em i)} 4083 MW field stars, which include the stars observed in the {\em Gaia}-ESO field samples by selecting the GES\_FLD keywords related to the field stars (GES\_MW for general Milky Way fields, GES\_MW\_BL for fields in the direction of the Galactic bulge, GES\_K2 for stars observed in  Kepler2 (K2) fields, GES\_CR for stars observed in CoRoT fields), and benchmark stars GES\_SD  and {\em ii)} 385 field stars that are non-members of open clusters (age$\,>\,$100~Myr), as described in Section~\ref{clusters}. 

We combined the two samples of field stars, performing a further selection on stellar parameter uncertainties: S/N > 20; $e T_{\rm eff} < 150$~K, $e{\rm log}\,g < 0.25$, $e {\rm [Fe/ H]}< 0.20,$ and $e \xi <$ 0.20 $km~ s^ {-1}$. We also applied a further cut in abundance errors, considering only those values that have an $e A(El) < 0.1$.
Our quality selection reduces the total number of MW field stars from 4468 to 3975, with  a median S/N$\sim$74.

The selected sample of 3975 stars is made up of 711 stars (18\%) with $ logg \leq 3.5 $ (giant stars) and 3264 stars (82\%) with $ logg> 3.5 $ (dwarf stars). Of these 3975 stars, 
%we have 3888 with R$_{\rm GC}$, of which 722 
19\% are located in the the inner disc, 78\% are in the solar region and 3\% belong to the outer disc. %Fig.~\ref{KD_field} shows the sample in the Kiel diagram with the bivariate distribution using kernel density estimation. 
Due to the selection function adopted in the {\em Gaia}-ESO survey for UVES observations \citep[see][]{stonkute16}, the field population is dominated by stars at the main sequence turn-off (MSTO), while several giant stars are present that are non-members of open clusters.

\section{Relations between age, R$_{\rm GC}$, and abundance ratios in open clusters}
\label{sec_multivariate}
\subsection{Age versus [El/Fe] relations}

The relation between the abundances of neutron-capture elements and stellar ages has been widely investigated \citep[e.g.][]{dorazi09, maiorca11, maiorca12, yong12, mishenina13a,mishenina13b, jacobson13, Battistini_2016, marsakov16, Nissen16, reddy17, delgadomena17, spina18, magrini18, tautvasiene21, casamiquela21, baratella21, zinn21, sales21}. 
However, most of the previous works were limited to the solar neighbourhood region, using both solar twins or star clusters. Exceptions are the works of \citet{magrini18}, which is based on a sample of open clusters from {\em Gaia}-ESO {\sc idr5}, about 20 clusters located at various galactocentric distances, and the very recent work of \citet{sales21}, in which a sample of 42 clusters from the APOGEE survey was used to investigate the spatial variation of the relation [Ce/$\alpha$] versus age, finding that is not the same across the Galactic disc. This is possibly due to the dependence of AGB yields on metallicity. 

Unlike previous studies, our sample offers the unique advantage of having a very large sample of OCs (62 objects) with homogeneously derived abundances of neutron-capture elements, covering a wide range of ages and galactocentric distances  (see Fig.~\ref{age_rgc_distribution}).
This allows us to study the relationships between ages and abundance ratios in a spatially resolved way and to derive them at various galactocentric distances. 
We divided our sample clusters in three galactocentric regions: 
an outer region, which includes 30 OCs located at $R_{gc}>9$ kpc; a central region, where our Sun is located, includes 20 OCs at $7 \leq R_{gc} \leq 9$ kpc; and an inner region, comprising 12 OCs at $R_{gc}<7$ kpc.

\begin{figure*}
  \resizebox{\hsize}{!}{\includegraphics{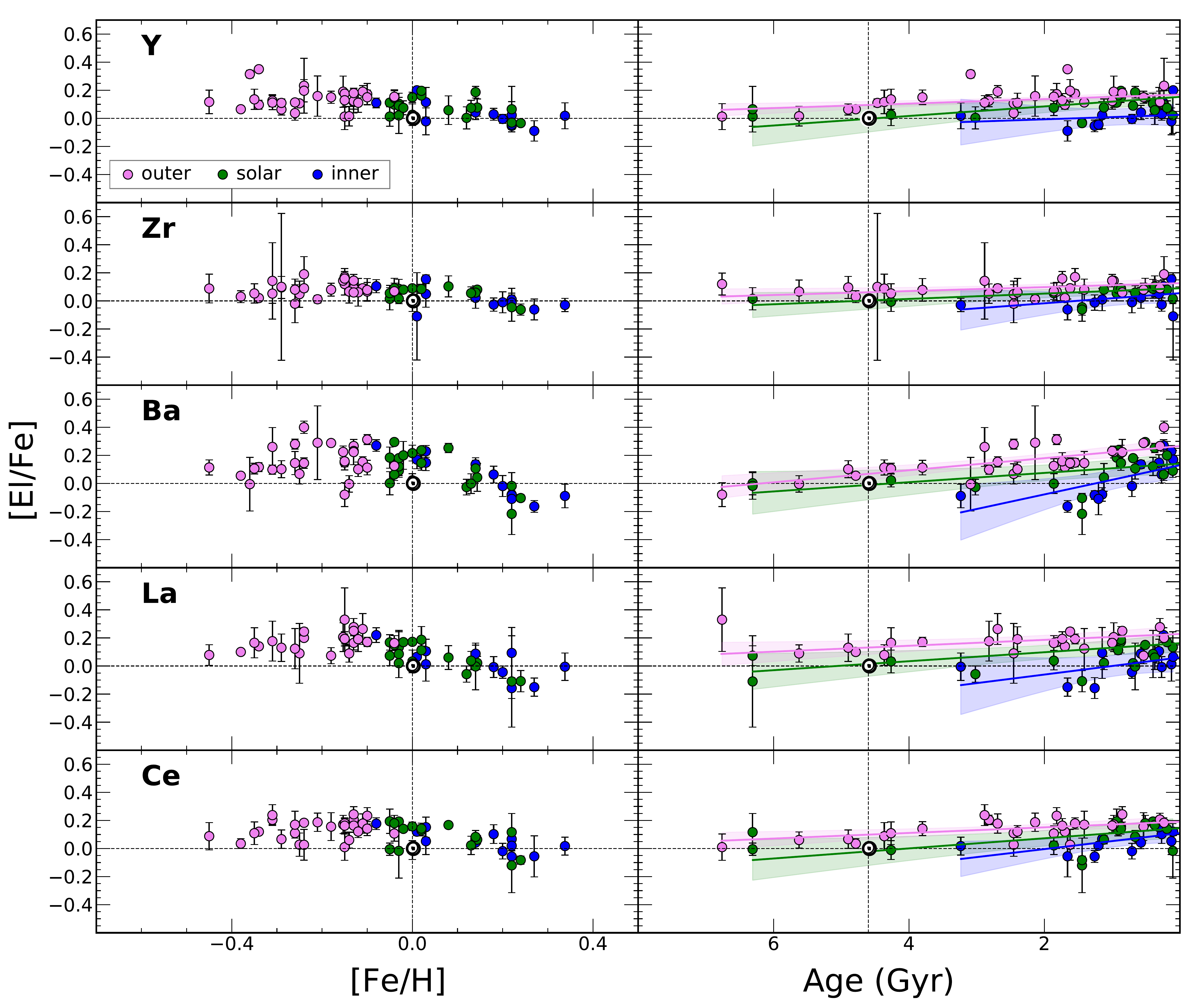}}
  \caption{[El/Fe] versus metallicity and age. In the left panels, we show [Fe/H] versus [El/Fe], and in the right panels we show age versus [El/Fe] for our sample of OCs divided into the three galactocentric regions. In blue, we show the OCs located in the inner disc, in green we show those located in the solar neighbourhood, and in pink we show the ones located in the outer disc.  In the right panels, the continuous lines are the univariate linear regressions (one for each radial region, colour-coded according to the corresponding sample of OCs), while the shaded regions are their confidence intervals.}
  \label{age_elfe1}
\end{figure*}

In Fig.~\ref{age_elfe1}, we show the variation of [El/Fe] as a function of [Fe/H] and of age. 
In the left panels, we observe important differences in the clusters belonging to the three regions mentioned above. The clusters in the outer disc are generally more metal poor than the clusters in the inner disc and in the solar neighbourhood, while the differences in [El/Fe] are less pronounced. Typically, outer disc clusters have [El/Fe]$>$0, while the inner disc ones have [El/Fe] solar or sub-solar. 
The results are more clear when looking at the right panels, in which [El/Fe] are shown as a function of the cluster age; for all the elements, [El/Fe] are underabundant for a given age in the inner disc with respect to those of the outermost regions. For all three regions, we observe an increasing trend of [El/Fe], confirming previous literature results. [Ba/Fe], as noted in the past \citep[e.g.][]{dorazi09, maiorca12, Mishenina15}, has the strongest upward trend. 

\begin{figure}
  \resizebox{\hsize}{!}{\includegraphics{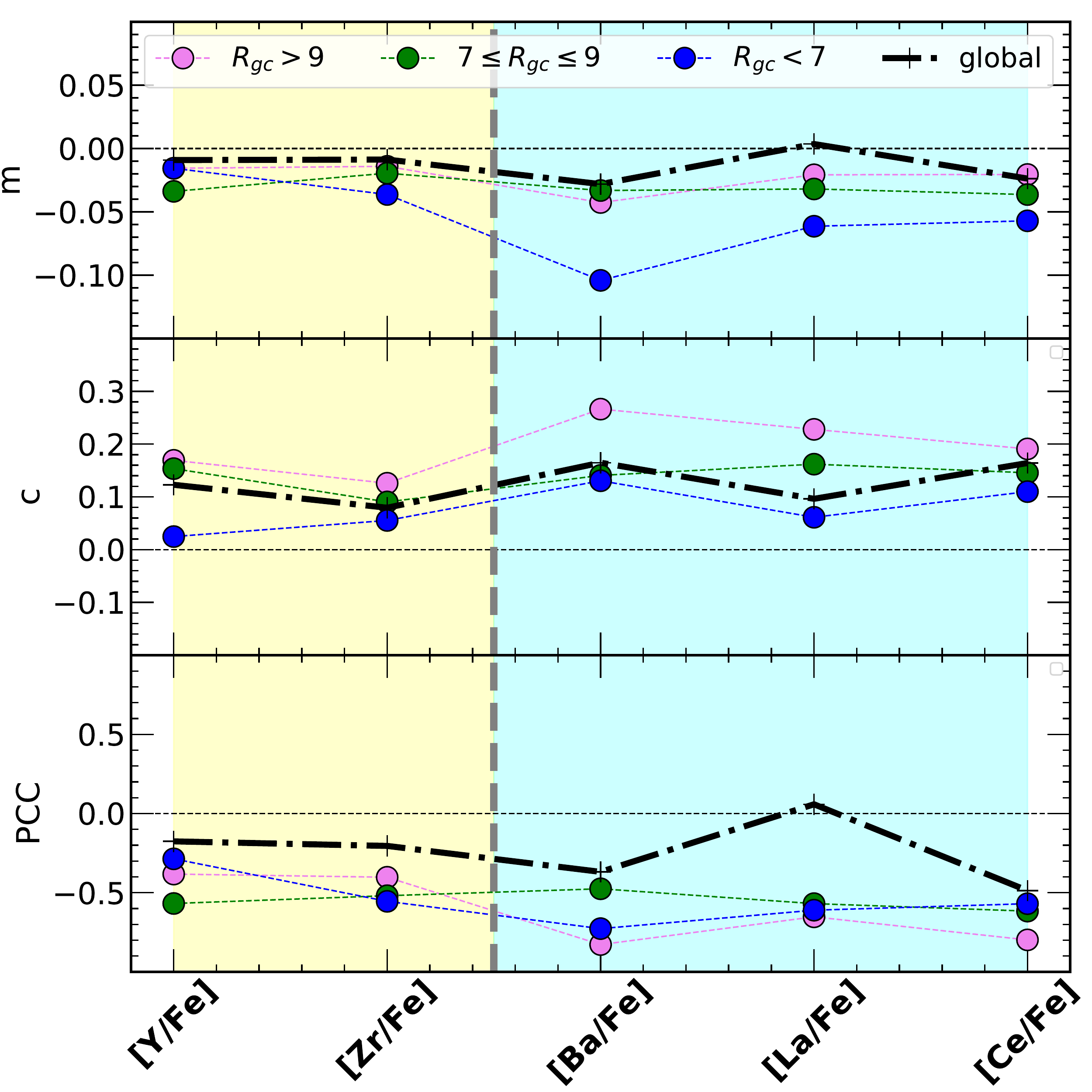}}
  \caption{Weighted regression coefficients for the age versus [El/Fe] relation of 62 OCs by R$_{\rm GC}$. The yellow area corresponds to the elements of the first $s$-process peak (Y and Zr), and the cyan area corresponds to those of the second peak (Ba, La and Ce). In the upper panel, we show the slope in dex~Gyr$^{-1}$ of the relations. In the central panel, we present the value of the intercept (dex), while in the bottom panel we see the correlation coefficient. In all panels, we use the following symbols: blue circles for the inner region, green circles for the solar region, pink circles for the outer region, and dashed black lines for the whole sample. }
  \label{fig:age_xfe_coeffs}
\end{figure}

In Fig.~\ref{fig:age_xfe_coeffs}, we show the results of the coefficients of the weighted regressions (WLS) in the three regions in the following form: [s/Fe]=$m$*Age+c. 
As a weight, we apply the expression:
$$
\frac{1}{(\frac{\sigma}{\sqrt{N}})^2,}
$$
where $\sigma$ is the standard deviation of the abundance ratios for the member stars of every cluster, and N is the number of member stars with values of the corresponding abundance ratio.
The coefficients are reported in Table~\ref{tab_coefficients} together with the Pearson correlation coefficients (PCCs). 

We note the different behaviour of the first-peak elements (Y and Zr) with respect to the second-peak ones (Ba, La, Ce), the former having lower intercept values than the latter.  
For all $s$-process-dominated elements, the slope of the regression is steeper in the inner disc than in the other two regions, while the value at the intercept, [El/Fe], is lower in that region (see Fig.~\ref{fig:age_xfe_coeffs}).
 In Table 2 of the appendix, we provide the coefficients of the weighted linear fits for the three radial regions. 

In the work of \citet{dorazi09}, the [Ba/Fe] ratio was found to dramatically increase at decreasing ages, with very high growth in very young stars.  As discussed in \citet{baratella21}, the higher enhancement of [Ba/Fe] with respect to [La/Fe] and [Ce/Fe] in young stars cannot be easily explained, either with non-local thermodynamic equilibrium (NLTE) effects or with stellar nucleosynthesis and chemical evolution models. To look for a plausible explanation, they explored different scenarios related to the formation and behaviour of spectral lines, from the dependence on the different ionisation stages and the sensitivity to the presence of magnetic fields and the effect of stellar activity. However, all these effects cannot fully explain the different behaviour of Ba in young stars. 
In our sample, we considered clusters older than 100 Myr; thus, we expect a relatively smaller difference between Ba, and the other elements of the second peak.  
However, as mentioned above, Figures~\ref{age_elfe1} and \ref{fig:age_xfe_coeffs} show that there are some differences among these elements, as already found in \citet{magrini18}. 
Following \citet{Mishenina15}, other possible explanations   are related to some extra production of Ba via an intermediate neutron-capture process, the so-called $i$-process, triggered by the mixing or ingestion of H in He-burning stellar layers \citep{cowan77, bertolli13}. 
To summarise, though the origin of the [Ba/Fe] increase in young stars does not have a complete theoretical explanation, the large slope of the [Ba/Fe] age relation makes Ba an excellent age proxy, at least for ages above 150~Myr \citep[see][]{spina20, baratella21}. %

\subsection{Age versus [s/$\alpha$] relations: Chemical clocks across the disc}

The idea of combining pairs of elements with different origins, particularly $s$ processes and $\alpha$ elements (or odd-Z elements such as Al), has taken shape in the last decade \citep[see, e.g.][]{tucci16, Nissen16, feltzing17, Fuhrmann17, slumstrup17, titarenko19, casali20}. The combination of the abundances of an $s$-process element with other elements with opposite behaviour, such as $\alpha$ elements, maximise their correlation with stellar age. Several relations have been established and calibrated in the solar neighbourhood \citep[see, e.g.][]{spina16, delgado19, jofre20, tautvasiene21}. Here, we aim to extend them in different regions of the Galactic disc using our sample of open clusters.  
For each neutron-capture peak, we selected the element with the highest percentage from an $s$ process in the Sun \citep[see, e.g.][]{arlandini99, Bisterzo14}: Y for the first peak and Ba for the second one. We combined them with several $\alpha$ elements, namely Mg, Si, Ca and Ti, and Al (which can be also considered an odd-Z element, but has a behaviour very similar to the other $\alpha$ elements).  

To find the best way to describe the relations between the age of the clusters and their chemical characteristics, we used multi-linear weighted regressions that take into account age, metallicity, and R$_{GC}$, in the following form: [s/$\alpha$] = $m_{1}\cdot$Age + $m_{2} \cdot$ R$_{\rm GC}$ + $m_{3} \cdot$ [Fe/H] + c. 
A similar approach was used in the work of \citet{casali20}, who took into account [Fe/H] in their sample of solar-like stars. 
We computed both a global regression, considering all clusters in the sample, and individual regressions for each radial region. For all regressions, we adopted the same weight system, as described in Section~4.1.
In Table \ref{tab:age_s_alpha_MLR_3_vbles}, we report the coefficients of the weighted multivariate regressions for the three radial regions, and for the global sample including all clusters at all R$_{\rm GC}$. 

\begin{figure*}
  \resizebox{\hsize}{!}{\includegraphics{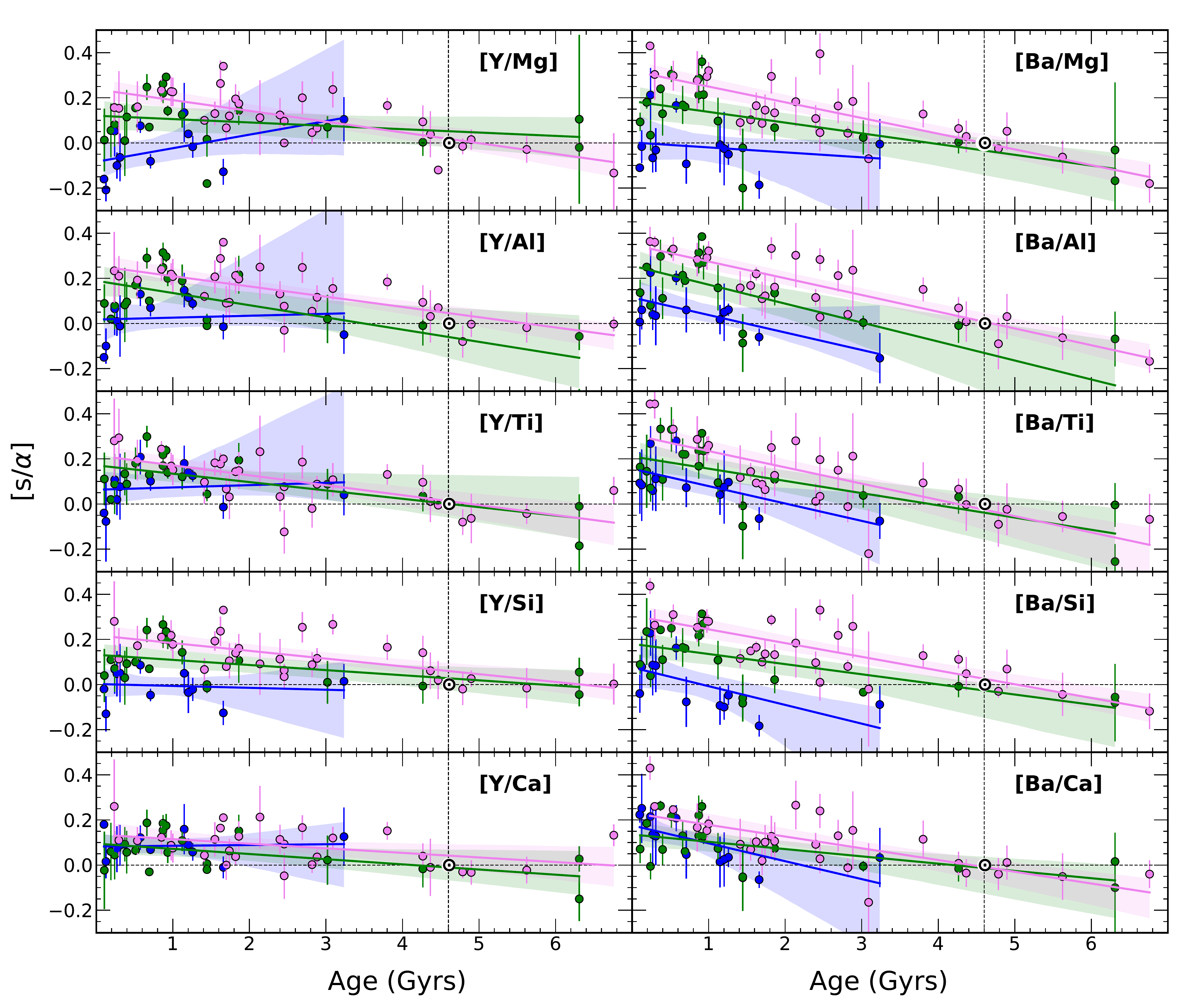}}
  \caption{Age versus [s/$\alpha$] for our OC sample divided among the three R$_{\rm GC}$ regions. The regression curves with their confidence intervals (shaded regions) are shown in each panel. On the left side, we have the abundance ratios [Y/$\alpha$] versus age, and on the right side we have  [Ba/$\alpha$] versus age. The symbols and colours are the same as in Fig.~\ref{age_elfe1}.}
  \label{age_ls_alpha}
\end{figure*}

In Fig.~\ref{age_ls_alpha},  we plot [Y/$\alpha$] and [Ba/$\alpha$] versus cluster ages in the three regions of the Galactic disc, as defined above.  Here, and in the following sections, we simplify our workings by including Al in the group of $\alpha$ elements, and thus when referring to [s/$\alpha$], we also consider Al. 
In the left panels of Fig.~\ref{age_ls_alpha}, we find well differentiated behaviours; in the outer and central regions, we have  decreasing trends of [Y/$\alpha$] with increasing age, while in the inner region we observe a reverse trend. This was already noted, with a smaller sample size, in \citet{magrini21a}. 
In the right panels of Fig.~\ref{age_ls_alpha}, the abundance ratios containing Ba and an $\alpha$ element show a decreasing trend with increasing age in all three regions. In addition, the innermost region shows a stronger trend than the others, with a steeper slope. 

\begin{figure}
  \resizebox{\hsize}{!}{\includegraphics{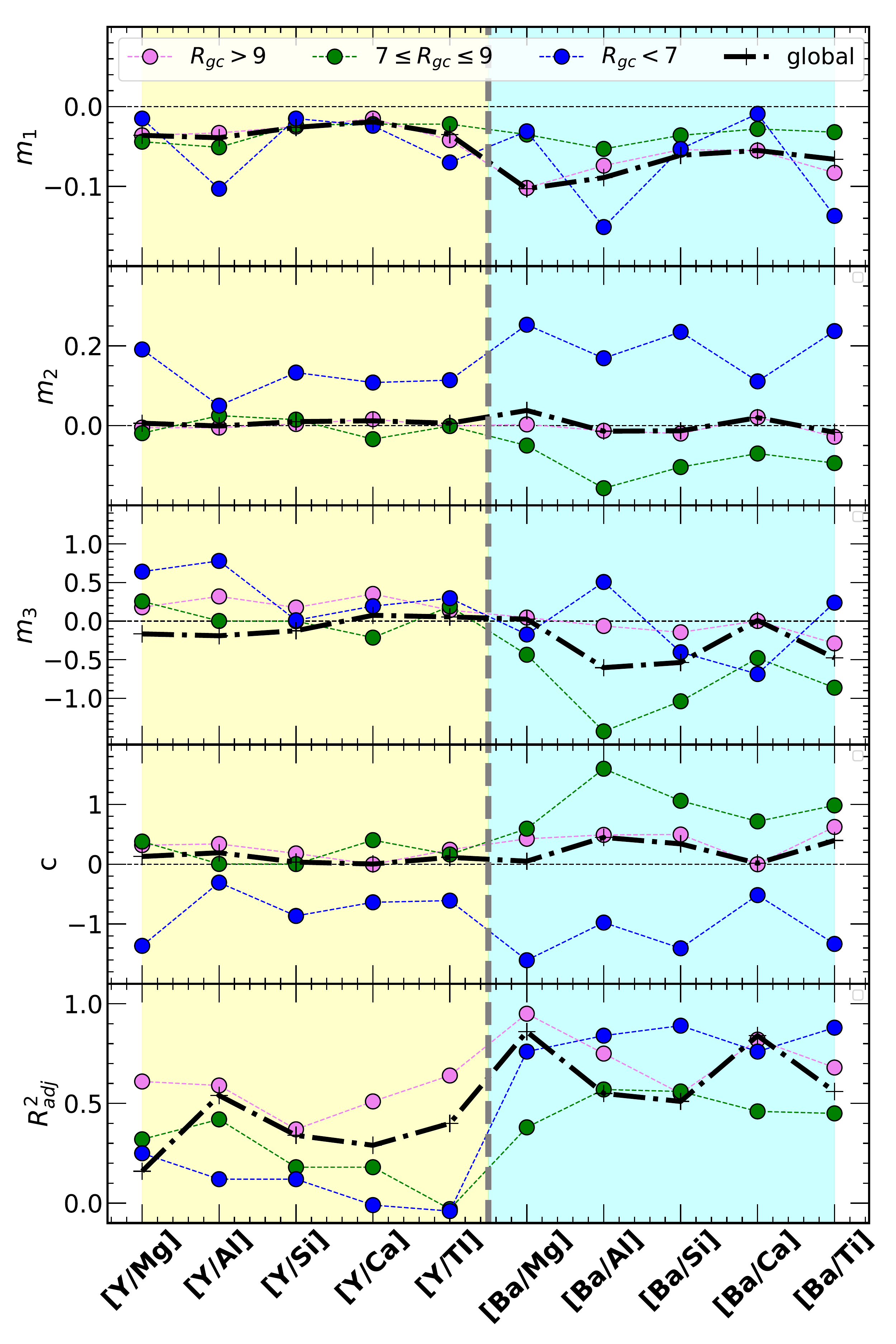}}
  \caption{Regression coefficients for the [s-/$\alpha$] = $m_{1}\cdot$age + $m_{2} \cdot$ R$_{\rm GC}$ + $m_{3} \cdot$ [Fe/H] + c relation. We show $m_{1 }$in the
upper panel, $m_{2}$ in the second one, $m_{3 }$ in the third one, and $c$ in
the fourth one.  In the bottom panel, we present the correlation coefficients. In all panels, we use the following symbols: blue circles for the inner region, green circles for the solar region, pink circles for the outer region, and dashed black lines for the whole sample. The yellow region highlights the abundance ratios with Y, and the cyan region shows them with Ba. }
  \label{age_ls_alpha_coeffs_mlr_3_vbles}
\end{figure}

The coefficients ($m1$, $m2$, $m3$ and $c$),  and the correlation coefficients are also shown in Fig.~\ref{age_ls_alpha_coeffs_mlr_3_vbles}, in which we separate the abundance ratios with Y  and with Ba, for each radial region  and for the whole sample.
It is interesting to notice that the region that differs the most from the others is the inner region, while that around the Sun and the outer region are very similar, and even similar to the global relationship. 
Therefore, the multivariate regression computed with the whole sample of clusters might still be a good approximation for the solar neighbourhood and the outer region, while it fails to reproduce the inner disc region \citep[cf.][]{casali20}.

%%%%%%%%%%%%%%%%%%%%%%%%%%%%%%%%%%%%%%%%%%%%%%%%%%%%%%%%%

%%%%%%%%%%%%%%%%%%%%%%%%%%%%%%%%%%%%%%%%%%%%%%%%%%%%%%%%%%

\subsection{Role of migration in open clusters}

\label{Sec_migration}

Radial migration plays an important role in the redistribution of stellar populations, particularly the older ones, in our Galaxy. What weight migration holds in shaping the spatial distribution of more massive populations such as clusters, with respect to single stars, is not yet settled \citep[see, e.g.][]{anders17, chen20, zhang21, netopil21}. 
To estimate the possible effect of radial migration in our relationships, we calculated the orbits of our cluster sample using the {\sc galpy} code, with the axis-symmetric potential {\sc MWPotential2014} \citep{Bovy_2015}.
We adopted the guiding radius \citep[R$_g$, defined as the average between the minimum and maximum radius, see, e.g.][]{halle15} instead of their present time R$_{\rm GC}$ to recompute the relations between age and abundance ratios, associating the clusters with the three radial regions on the basis of their  R$_g$. Adopting R$_g$ can indeed mitigate the effect of blurring due to epicyclic oscillations around the guiding radius \citep{SB09}, while it cannot overcome  the migrating effect of churning, that is the change of R$_g$ due to interactions with a lasting non-axisymmetric pattern such as long-lived spiral arms or long-lived bars \citep{SB02, BT08}.
Using R$_g$,  we found a new redistribution in the three regions:  eight clusters might be visitors in their respective, initially assigned regions. The  most affected region is the inner disc, which is repopulated with seven clusters coming mostly from the central region of the disc, with the exception of one cluster coming from the outer region (ESO 92 05). The outer disc is less affected by redistribution. The other seven clusters that move from the solar region to the inner region are NGC6791, Berkeley 44, NGC6802, NGC4815, Trumpler 20, NGC4337, and Collinder 261.
Among them, NGC6791, which is both old and metal rich, is known to have a high probability of being a migrator \citep[cf.][]{Jilkova_2012, netopil21}.  
Taking into account the effect of blurring, the regions are redistributed to contain 20 OCs in the inner disc instead of 7; 13 in the solar region instead of 20; and 29 in the outer region instead of 30.

 We recalculated the relations adopting R$_g$ instead of R$_{\rm GC}$. 
 As expected, only regressions for the inner and solar regions change when replacing R$_{GC}$ with R$_g$. There is not an improvement in the quality of the fits, with a decrease in the correlation coefficients in the inner disc and in the solar neighbourhood (about 0.1 lower than those obtained with  R$_{\rm GC}$).
In the next sections, we use the relationships obtained with the current galactocentric position.

 \subsection{Comparison with literature results} 

\begin{figure*}
  \resizebox{\hsize}{!}{\includegraphics{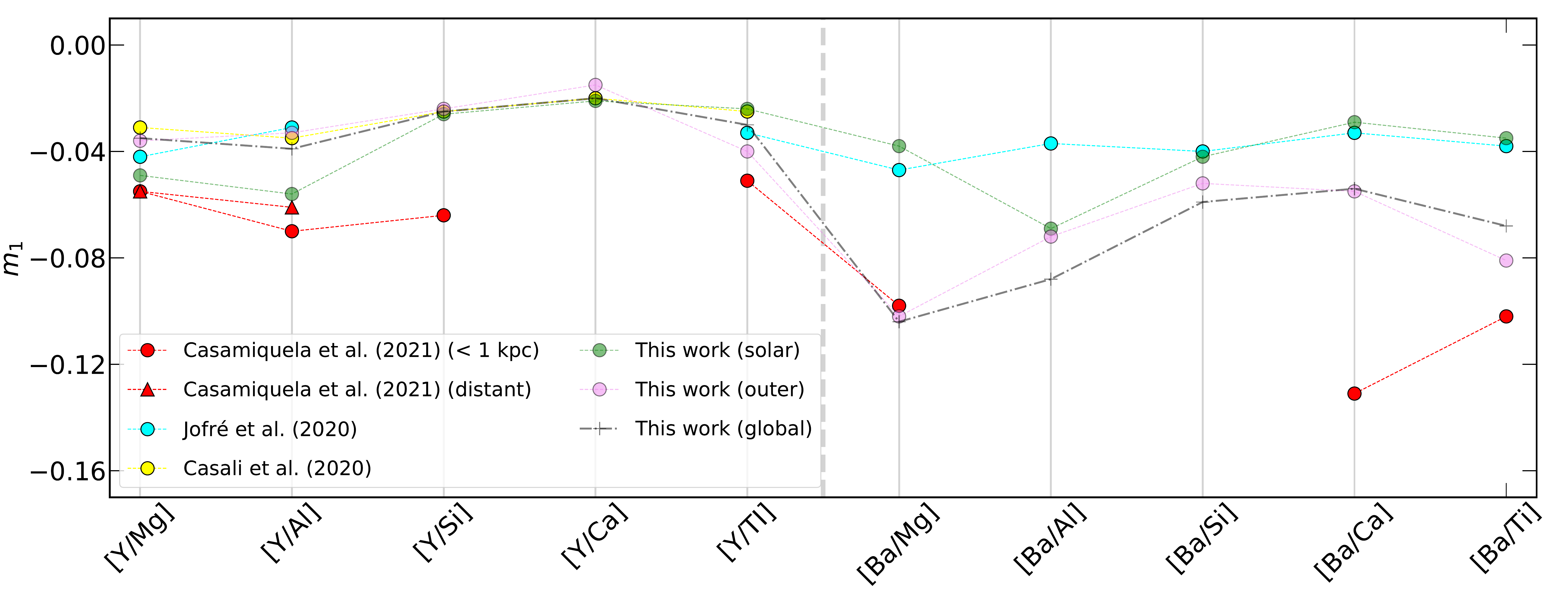}}
  \caption{Comparison of first independent variable ($m_1$ $\cdotp$ age) of our stellar dating relations with those of \citet{casali20} (yellow) for solar-like stars (which were tested in open clusters), \citet{jofre20} for a sample of 80 solar twins (cyan), and those of \citet{casamiquela21} for a sample of 47 open clusters (red).}
  \label{fig:comparison}
\end{figure*}

In this section, we compare our results with three recent literature works \citep{casali20, jofre20, casamiquela21}. In Fig.~\ref{fig:comparison}, we show the coefficient $m_1$ related to {\em age} and computed for the samples of open clusters in the outer and solar regions  with the results from \citet{casali20} (yellow dots) for a sample of solar-like stars (and later applied to open clusters), those of \citet{casamiquela21} for two samples of open clusters (divided and closer than 1 kpc to the Sun and more distant; red dots and triangles respectively), and  those of
\citet{jofre20} for a sample of 80 solar twins (cyan dots in Fig.~\ref{fig:comparison}).
The samples of solar twins or solar-like stars are to be compared with our sample of clusters in the solar region, while our outer disc sample can be compared with the outermost sample of \citet{casamiquela21}. 
The general agreement is very good, and our slopes for the solar region sample agree well with the solar-like stars from different authors. 
The largest differences are seen in a sample of 
\citet{casamiquela21} for [Ba/Ca] and [Ba/Ti], which might be related to the inclusion of younger clusters in their sample.

\section{Application to cluster stars}
\label{sec_ages_clusters}
\subsection{Global ages of open clusters} 

\begin{figure*}
  \resizebox{\hsize}{!}{\includegraphics{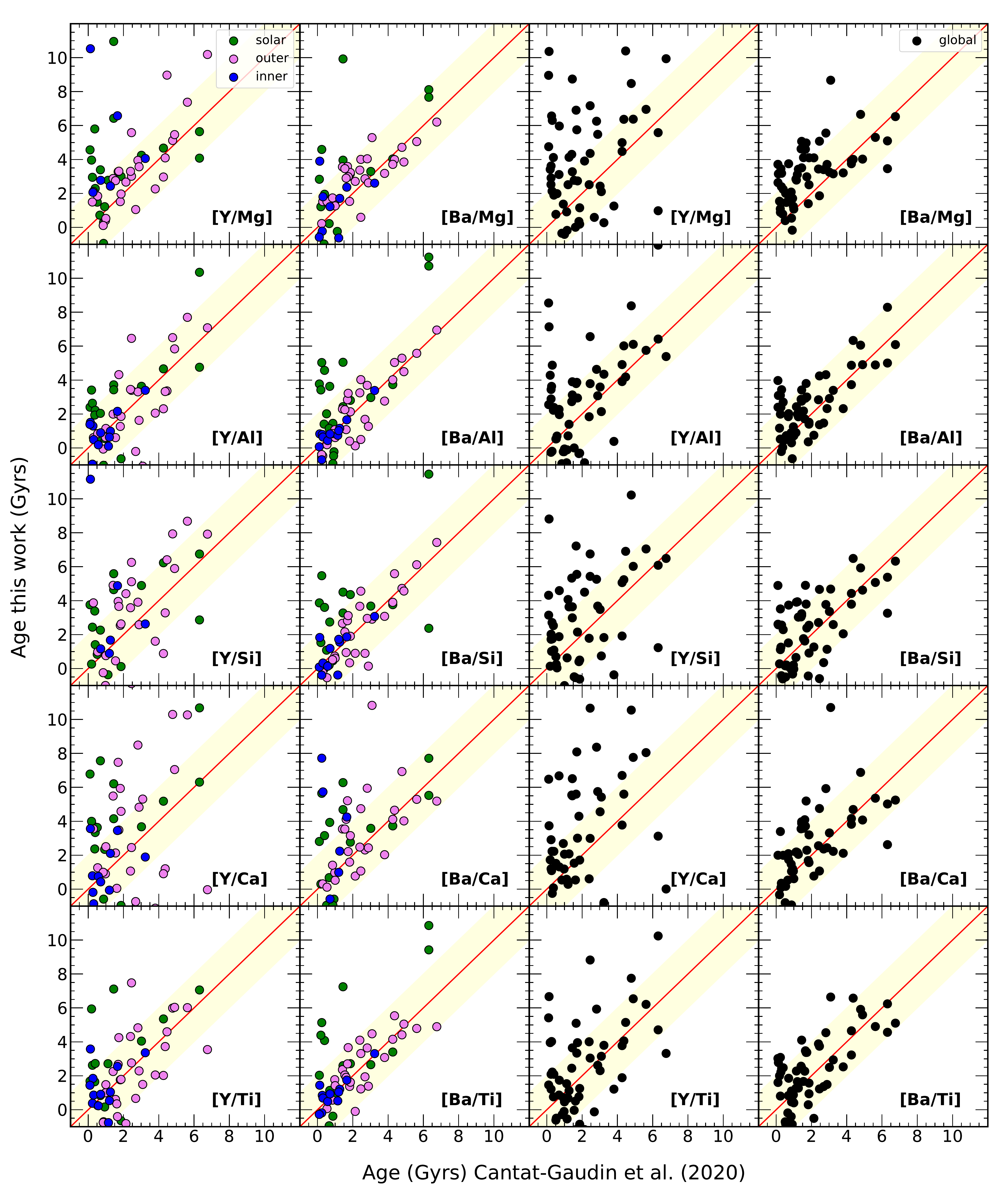}}
  \caption{Ages computed with our relations  versus ages from isochrone fitting \citep{CG20}. In the left panels, we show the ages computed  with the relations obtained for each radial region. The clusters are colour-coded by R$_{\rm GC}$. In the two right panels, we show the ages computed with global relations versus the ages from \citet{CG20}.   The red continuous line is the bisector and indicates the 1-to-1 relation. The yellow shaded regions indicate the clusters with ages within 2~Gyr from their age from the isochrone fitting. }
  \label{ageiso_agecheclock}
\end{figure*}

The first step in verifying the validity of our relations and their ability to provide a reliable estimate for the stellar ages is to reapply the same relations to clusters, by comparing the input ages from isochrone fitting with those obtained from the ten considered chemical clocks.
The comparison between the two ages is shown in Fig.~\ref{ageiso_agecheclock}.
The cluster ages, for each range of R$_{\rm GC}$, are computed with the corresponding radial relations or with the global relations (throughout the entire R$_{\rm GC}$ interval). 
In the panels in each row, we show the results for a given chemical clock. 
There are three main aspects to note: {\em i)} the scatter increases for younger ages for almost all relations, {\em ii)} some relations allow us to recover the input ages with greater accuracy (closer to the 1-to-1 relation) and precision (less scatter) than the others. 
We defined our accuracy as the mean average difference between input and output ages obtained from our relations (which is indeed a sort of mean bias with respect to the reference values), while the precision (which can be considered as the scatter or internal error calibration) is the standard deviation of the mean difference; {\em iii)} the global relations generally produce worst results, especially for the younger ages (which is expected, since younger clusters dominate the inner region, for which the global relations do not work). 
The accuracy and precision, as defined above, for all radial bins, including the global one, and for all the considered abundance ratios are shown in Table~\ref{tab:accuracy}.

The best set of relations, in terms of both accuracy, precision, and recovering also of the younger ages, are those involving [Ba/Al], which has  some of the highest correlation coefficients in the three galactocentric regions. 
As can be seen in Table \ref{tab:age_s_alpha_MLR_3_vbles}, the adjusted coefficients of determination using the [Ba/Al] ratio are 0.75, 0.57, and 0.84 for the outer, solar, and inner regions, respectively. 
On the other hand, the global relation has a lower correlation coefficient of 0.55. 

As shown in Table~\ref{tab:accuracy}, the accuracy obtained with the [Ba/Al] relation is 0.1 Gyr in the outer region, $-1.4$ Gyr in the solar region, and $-0.0$ Gyr in the inner region, with precisions of 0.9, 2.4, and 0.4 Gyr, respectively. 
However, if we exclude in the solar region, the three most discrepant clusters,  NGC6971, Berkeley 44, and Collinder 261, which are likely subject to migration (see Section~\ref{Sec_migration}), the accuracy improves, decreasing to $\sim$ $-0.9$ Gyr, and the precision becomes slightly lower at 2.3 Gyr. These numbers give us a first estimate of the kind of uncertainties to which ages measured with chemical clocks are subject.

\begin{table}
\begin{threeparttable}[hb]
    \caption{Accuracy and precision of our relations in recovering the ages of clusters in each of the defined regions and globally.}              % title of Table
    \label{tab:accuracy}      % is used to refer this table in the text
    \centering      % used for centering table
    \begin{tabular}{lrr}      
    \hline\hline % inserts double horizontal lines
        & R$_{\rm GC}$ > 9&  \\
    \hline % inserts single horizontal line
    ratio & accuracy (Gyr) & precision (Gyr) \\    % table heading
    \hline
  $[\ion{Y}{ii}/\ion{Mg}{i}]$ & $-$0.2 & 2.1 \\
  $[\ion{Y}{ii}/\ion{Al}{i}]$ & 0.6 & 2.2 \\
  $[\ion{Y}{ii}/\ion{Si}{i}]$ & 0.1 & 2.9 \\
  $[\ion{Y}{ii}/\ion{Ca}{i}]$ & $-$0.1 & 4.7 \\
  $[\ion{Y}{ii}/\ion{Ti}{i}]$ & 0.2 & 1.7 \\
    \hline
  $[\ion{Ba}{ii}/\ion{Mg}{i}]$ & $-$0.5 & 1.0 \\
  $[\ion{Ba}{ii}/\ion{Al}{i}]$ & 0.1 & 0.9 \\
  $[\ion{Ba}{ii}/\ion{Si}{i}]$ & 0.3 & 1.3 \\
  $[\ion{Ba}{ii}/\ion{Ca}{i}]$ & $-$0.5 & 2.0 \\
  $[\ion{Ba}{ii}/\ion{Ti}{i}]$ & 0.1 & 1.1 \\
    \hline
         & 7 < R$_{\rm GC}$ < 9 (kpc)$^\ast$ & \\
    \hline
  $[\ion{Y}{ii}/\ion{Mg}{i}]$ & $-$1.6 ($-$1.6) & 2.8 (2.1)\\
  $[\ion{Y}{ii}/\ion{Al}{i}]$ & $-$0.3 ($-$0.1) & 2.1 (1.9)\\
  $[\ion{Y}{ii}/\ion{Si}{i}]$ & 0.5 (0.6) & 3.4 (3.5)\\
  $[\ion{Y}{ii}/\ion{Ca}{i}]$ & $-$1.3 ($-$0.9) & 3.3 (3.3)\\
  $[\ion{Y}{ii}/\ion{Ti}{i}]$ & $-$0.7 ($-$0.1) & 3.8 (3.1)\\
    \hline
  $[\ion{Ba}{ii}/\ion{Mg}{i}]$ & $-$0.3 (0.3) & 3 (2.4)\\
  $[\ion{Ba}{ii}/\ion{Al}{i}]$ & $-$1.4 ($-$0.9) & 2.4 (2.3)\\
  $[\ion{Ba}{ii}/\ion{Si}{i}]$ & $-$0.4 ($-$0.3) & 2.9 (2.6)\\
  $[\ion{Ba}{ii}/\ion{Ca}{i}]$ & $-$0.3 ($-$0.1) & 2.6 (2.6)\\
  $[\ion{Ba}{ii}/\ion{Ti}{i}]$ & $-$0.6 (0.1) & 2.8 (2.5)\\
    \hline
        & R$_{\rm GC}$ < 7 (kpc)  &  \\
    \hline
  $[\ion{Y}{ii}/\ion{Mg}{i}]$ & 2.0 & 7.0\\
  $[\ion{Y}{ii}/\ion{Al}{i}]$ & $-$0.1 & 0.8\\
  $[\ion{Y}{ii}/\ion{Si}{i}]$ & 1.2 & 4.9\\
  $[\ion{Y}{ii}/\ion{Ca}{i}]$ & 0.9 & 2.9\\
  $[\ion{Y}{ii}/\ion{Ti}{i}]$ & $-$0.4 & 1.3\\
    \hline
  $[\ion{Ba}{ii}/\ion{Mg}{i}]$ & 0.7 & 2.3\\
  $[\ion{Ba}{ii}/\ion{Al}{i}]$ & $-$0.0 & 0.4\\
  $[\ion{Ba}{ii}/\ion{Si}{i}]$ & 0.2 & 0.9\\
  $[\ion{Ba}{ii}/\ion{Ca}{i}]$ & 2.2 & 5.5\\
  $[\ion{Ba}{ii}/\ion{Ti}{i}]$ & $-$0.1 & 0.4\\
    \hline
         & 6 < R$_{\rm GC}$ < 20 (kpc) &  \\
    \hline
  $[\ion{Y}{ii}/\ion{Mg}{i}]$ &$-$1.3   & 3.1\\
  $[\ion{Y}{ii}/\ion{Al}{i}]$ & $-$0.3 & 2.7\\
  $[\ion{Y}{ii}/\ion{Si}{i}]$ & 0.0 & 3.2\\
  $[\ion{Y}{ii}/\ion{Ca}{i}]$ & $-$0.3 & 3.7\\
  $[\ion{Y}{ii}/\ion{Ti}{i}]$ & $-$0.1 & 2.4\\
    \hline
  $[\ion{Ba}{ii}/\ion{Mg}{i}]$ & $-$1.2 & 1.5\\
  $[\ion{Ba}{ii}/\ion{Al}{i}]$ & $-$0.6 & 1.3\\
  $[\ion{Ba}{ii}/\ion{Si}{i}]$ & $-$0.3 & 1.7\\
  $[\ion{Ba}{ii}/\ion{Ca}{i}]$ & $-$0.2 & 1.8\\
  $[\ion{Ba}{ii}/\ion{Ti}{i}]$ & $-$0.2 & 1.5\\
    \hline
    \end{tabular}
    \begin{tablenotes}
        \item[$^\ast$] Results after removing discrepant OCs are included in parentheses.
    \end{tablenotes}
\end{threeparttable}
\end{table}

\subsection{Ages of individual member stars} 
\begin{figure*}
\centering
  \includegraphics[scale=0.45]{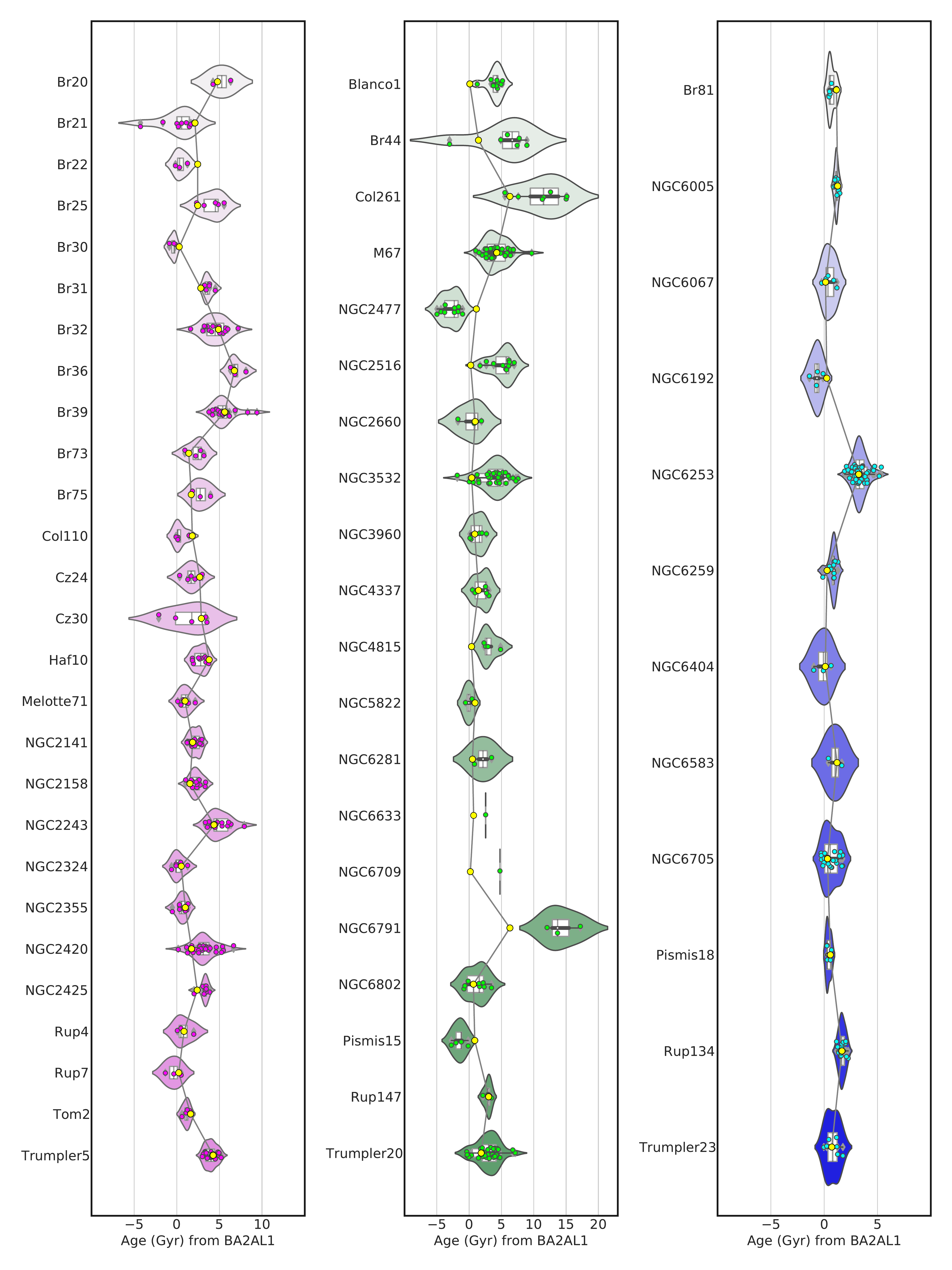}
  \caption{Violin plots of the ages of each stellar member of the clusters calculated using the relations inferred from [Ba/Al] for each of the three regions. Each of the coloured dots represents the age of each member star calculated individually, while the yellow dot is the age from \citet{CG20} for each of the clusters. The thick white bar in the center represents the interquartile range and the thin grey line represents the 1.5$\times$IQR, also showing a kernel density estimation with the distribution shape of the data. 
  The left panel shows the clusters in the outer disc, the central panel those in the solar region, and the right one the inner disc clusters. 
  }
  \label{violin_member_stars}
\end{figure*}

Another interesting test is to compare the age of each cluster member with the age obtained from one of our best relations. We selected the ages computed with the relations based on the abundance ratio [Ba/Al] as an example. This comparison allowed us to make a more realistic estimate of the uncertainties when the relation is applied to the field stars. 
Within the same cluster, there can be  considerable variations in abundance ratios, which are linked to the quality of the measurements and some peculiar enrichment in Ba and in other $s$-process elements. This results in variations between the estimated ages of members of the same clusters, which are expected to be coeval. 
In Figure \ref{violin_member_stars}, we show the violin plots of the ages of each stellar member of the clusters calculated using the relations between [Ba/Al] and age, derived independently in the three radial regions. We compare the individual ages with the literature ages from \citet{CG20}.  
In the outer and inner discs, the isochrone fitting ages for almost all clusters falls within the interquartile range of the ages calculated with our relations. However, in the solar region, this does not happen for some clusters, such as NGC6791, Collinder 261, NGC4815, or Berkeley 44. As discussed in Section~\ref{Sec_migration}, this supports the high probability of these clusters being subject to migration. We suppose that they were born in the inner disc, as evidenced by their $R_{g}$. We then did a further test, calculating their ages with the relation obtained for the inner disc. We have an improvement for NGC6791 and Berkeley 44, while the age of Collinder 261 remains discrepant. We also have to consider that the age given for NGC6791 by \citet{CG20} is a lower limit, and there are many works that give higher ages for this cluster \citep[e.g.][]{Brogaard_2011,Brogaard_2012, Brogaard_2021}; thus, they are similar to the age obtained with our relations. Furthermore, the above seems to be supported by the fact that other clusters in the solar region that appear to deviate in Figure \ref{violin_member_stars} (e.g. NGC6709, Pismis15, NGC5822, NGC6633, NGC2516) have $R_{\rm GC}$ at the edge with the inner region (between 7-7.5 kpc).

\section{ Application to field stars}
\label{ages_fiels}

The last step of our analysis was to verify the applicability of the relations  to the field stars. To do this, we needed to select a sample of field stars with similar characteristics to those of the  clusters with which we calibrated the relations. They must therefore be thin-disc stars, in approximately the same [Fe/H] range as our open cluster sample, and their ages must not exceed 7 Gyr.
The age constraint is the most difficult to obtain, because we know that the determination of the age of field stars with isochrone fitting is extremely uncertain. We thus relaxed this constraint, but we have to take it into account when we measure very old ages for stars with our relations. Those ages fall outside the limits for which they were calibrated. 
We used a chemical criterion to discriminate stars potentially outside the thin disc, based on the separation proposed by \citet{Adibekyan_2012}, obtaining a sample of about 2600 thin disc stars.

\begin{figure*}
  \resizebox{\hsize}{!}{\includegraphics{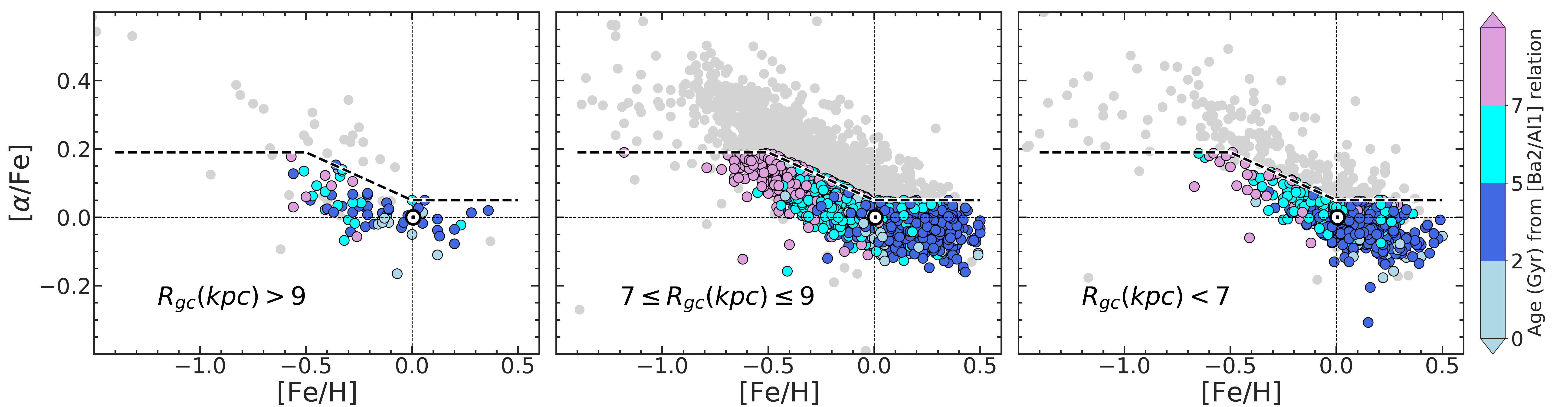}}
  \caption{[Fe/H] versus [$\alpha$/Fe] for field stars in the three radial regions coloured with the inferred ages using the [Ba/Al] relations computed for each radial region. The dashed black line is the dividing line given by \citet{Adibekyan_2012}. In the background (grey), we show the field stars excluded from our selection, which are presumably thick-disc and halo stars.}
  \label{field_from_ba2al1_by_Rgc}
\end{figure*}

As in the previous test, we show an illustrative case here, presenting the results from the relationship with the best correlation, accuracy, and precision, that is the one between age and [Ba/Al]. In Fig.~\ref{field_from_ba2al1_by_Rgc}, we show [$\alpha$/Fe]\footnote{[$\alpha$/Fe] is computed as ([Mg/Fe]+[Si/Fe]+[Ca/Fe]+[Ti/Fe])/4} versus [Fe/H],   separating the three radial regions. 
Our sample is dominated by the solar neighbourhood stars by construction \citep{stonkute16}, and the inner and outer bins are under-sampled. 
The results of the age distribution in [$\alpha$/Fe]-[Fe/H] plane agree with the expectation for the age distribution in the thin disc \citep[see, e.g.][]{haywood13, hayden15, buder19, casali19, casali20}. The oldest stars are those with a higher [$\alpha$/Fe] ratio, while the youngest have solar or slightly sub-solar ratios. There are few $\alpha$-enhanced stars at high metallicity and of intermediate age. 
 We remind the reader that these are only probabilistic indications of age and not real measurements. For field stars, the scatter in abundance ratios of age tracers at a given age (calculated with isochrones) is about 0.2 dex. 
This makes the age determination with chemical clocks only approximate and of purely statistical value.

\section{Summary and conclusions}
\label{sec_discussion}

Here, we list some considerations and caveats on the limit of the relations providing ages based on chemical clocks (related to chemical evolution), such as the ones used in the present work.  
\begin{itemize}
    \item[i.] {\bf The role of star formation history (SFH)} --Such relations can only be applied to the same populations from which they were derived. Since the abundance ratios [$s$/$\alpha$] strongly depend on the SFH, we recommend only using the relations calibrated with open clusters for the thin disc field population, in the same metallicity interval sampled by the calibrators used. Ages older than those of the calibrators are a dangerous extrapolation. 
    \item[ii.] {\bf The role of migration} --Field stars can migrate. Although we saw that the effect on the slopes of our relations due to blurring on open clusters is limited, this effect might be larger for field stars, and it can be combined with an even stronger effect of churning. There are some empirical methods based on deriving, for each star, its birth radius using its current metallicity, R$_{\rm GC,}$ and an observed or modelled radial metallicity gradient \citep[see, e.g.][]{feltzing17, quillen}. These methods introduce other uncertainties depending on the choice of gradient and its time evolution, particularly for older stars.
    So, we point out that the ages derived with the relations of the present work do not take into account migration, and if the effect of migration is greater than the size of the radial bin under consideration, they may provide incorrect ages. This is especially true in the inner disc, where migration to the outer disc dominates, but it can also affect the outer disc since stellar migration can be bi-directional. 
    \item[iii] {\bf The non-uniqueness of the results} -In this paper, we provide ten different relationships: five based on Y and five based on Ba. Although the relationships are in general agreement, they do not provide the same age for each star. We preferred to leave the relationships separate so that, depending on which abundances are available, it might be possible to provide an age estimate. The simultaneous use of several indicators gives an idea of the range of possible ages. We remind the reader that our relations are not to be used for an individual measurement of the age of field stars, but to statistically infer the age distribution of a population or characterise a single star as young, intermediate-age, or old. 
\end{itemize}

At the moment, this is the best we can do. With our sample of clusters, we mapped the spatial variations of the age-chemical-clock relationships. We confirm that these are not unique and that they vary more in the inner disc. 
Although the cluster sample is the largest used so far, it has an inherent limitation due to the characteristic of cluster populations: it is a thin-disc population with young and intermediate ages. In particular, the inner disc population being subject to more destructive effects is limited to clusters younger than 3 Gyr. This is a general limit that is not related to our sample but is also visible in larger samples \citep{CG20}. 
The strong recommendation we repeat is not to extrapolate the relationships into unmapped intervals from the clusters, and not to apply it to populations other than the thin disc. 
In the future, samples combining asteroseismological and spectroscopic observations will help us to extend the relationships to older ages and to populations other than the thin disc.

%\section{Conclusions}
%\label{sec_conclusions}

In the present work, we provided a set of multiple weighted linear regressions in three variables ([Fe/H], R$_{\rm GC}$, and age) between abundance ratios, the so-called chemical clocks, and stellar ages, which were calibrated with an appropriate sample of open clusters, the ages and distances of which were homogeneously determined using {\em Gaia} {\sc dr2}. 
We  estimate the accuracy and precision of each relation in recovering the age of open clusters. 
Among the considered chemical clocks, [Ba/Al], and in general abundance ratios involving barium,  provide the best recovering factor. The relation between [Ba/Al] and age is also able to reproduce the ages of the individual member stars with a precision better than 2 Gyr. Considering that typical errors on the ages of giant stars by isochrone fitting can reach 100\%, the result is very encouraging. 
In the solar region, this relationship does not work for some clusters (e.g. NGC6791, Berkeley 44, NGC4815, and Collinder 261), which are likely  subjects of migration.
We applied our relations to a sample of thin-disc field stars, selected from the {\em Gaia}-ESO {\sc idr6} catalogue. This ensures that we have a homogeneous analysis with respect to the clusters used to calibrate the relationships. 
Since the sample of field stars contains both dwarfs and giants, we used a solar scale based on giants and dwarfs of M67 to avoid any offset between the two samples. 
Using chemical information, we separated the field sample in Galactic populations: thin disc, thick disc, and halo. We only considered thin-disc stars in each radial region, in approximately the same [Fe/H] range of the open clusters used to calibrate the relations. 
We computed their ages: the distribution of ages in the [$\alpha$/Fe] versus [Fe/H] plane show the expected behaviour with youngest stars at high [Fe/H] and low [$\alpha$/Fe], and the oldest ones at low [Fe/H] and high [$\alpha$/Fe]. 
Finally, we discuss the limits of our method and give some important recommendations on the use of ages from chemical clocks based on chemical evolution. 
We state that future improvements can be made through the use of large asteroseismic samples combined with high-resolution spectroscopy, which will allow us to obtain calibrators in the old age regime, which are currently missing. 
In a forthcoming paper, we aim also to discuss the origin of the radial variations of the [s/$\alpha$] ratios in detail and in terms of chemical evolution and stellar nucleosynthesis.

%\begin{figure*}
%  \resizebox{\hsize}{!}{\includegraphics{AgeComparison_63OC.png}}
%  \caption{Comparison between literature age and age from chemical clocks for 63 open clusters.}
%  \label{agecomparison}
%\end{figure*}

\begin{acknowledgements}
We thank an anonymous referee for her/his careful reading and valuable comments to the first version of the manuscript.
Based on data products from observations made with ESO Telescopes at the La Silla Paranal Observatory under programme ID 188.B-3002. These data products have been processed by the Cambridge Astronomy Survey Unit (CASU) at the Institute of Astronomy, University of Cambridge, and by the FLAMES/UVES reduction team at INAF/Osservatorio Astrofisico di Arcetri. These data have been obtained from the Gaia-ESO Survey Data Archive, prepared and hosted by the Wide Field Astronomy Unit, Institute for Astronomy, University of Edinburgh, which is funded by the UK Science and Technology Facilities Council.
This work was partly supported by the European Union FP7 programme through ERC grant number 320360 and by the Leverhulme Trust through grant RPG-2012-541. We acknowledge the support from INAF and Ministero dell' Istruzione, dell' Universit\`a' e della Ricerca (MIUR) in the form of the grant "Premiale VLT 2012" and "Premiale 2016 MITiC". The results presented here benefit from discussions held during the Gaia-ESO workshops and conferences supported by the ESF (European Science Foundation) through the GREAT Research Network Programme.
This work has made use of data from the European Space Agency (ESA) mission {\it Gaia} (\url{https://www.cosmos.esa.int/gaia}), processed by the {\it Gaia} Data Processing and Analysis Consortium (DPAC, \url{https://www.cosmos.esa.int/web/gaia/dpac/consortium}). Funding for the DPAC has been provided by national institutions, in particular the institutions participating in the {\it Gaia} Multilateral Agreement. CVV and LM thank the COST Action CA18104: MW-Gaia. TB was funded by grant No. 2018-04857 from The Swedish Research Council. F.J.E. acknowledges financial support by the spanish grant PGC2018-101950-B-I00 and MDM-2017-0737 at Centro de Astrobiología (CSIC-INTA), Unidad de Excelencia María de Maeztu, and from the European Union’s Horizon 2020 research and innovation programme under Grant Agreement no. 824064 through the ESCAPE - The European Science Cluster of Astronomy $\&$ Particle Physics ESFRI Research Infrastructures project. CVV is especially grateful to Andrés Moya Bedón (University of Valencia) for his support and help.
\end{acknowledgements}

\bibliographystyle{aa} % style aa.bst
\bibliography{42937corr} % your references Yourfile.bib

%-----------------------------------------------------------------
\begin{appendix}
\section{Additional material}

%%%%%%%%%%%%%%% outliers %%%%%%%%%%%%%%%

\begin{figure*}
\begin{subfigure}{\textwidth}
  \centering
  % include first image
  \includegraphics[scale=0.35]{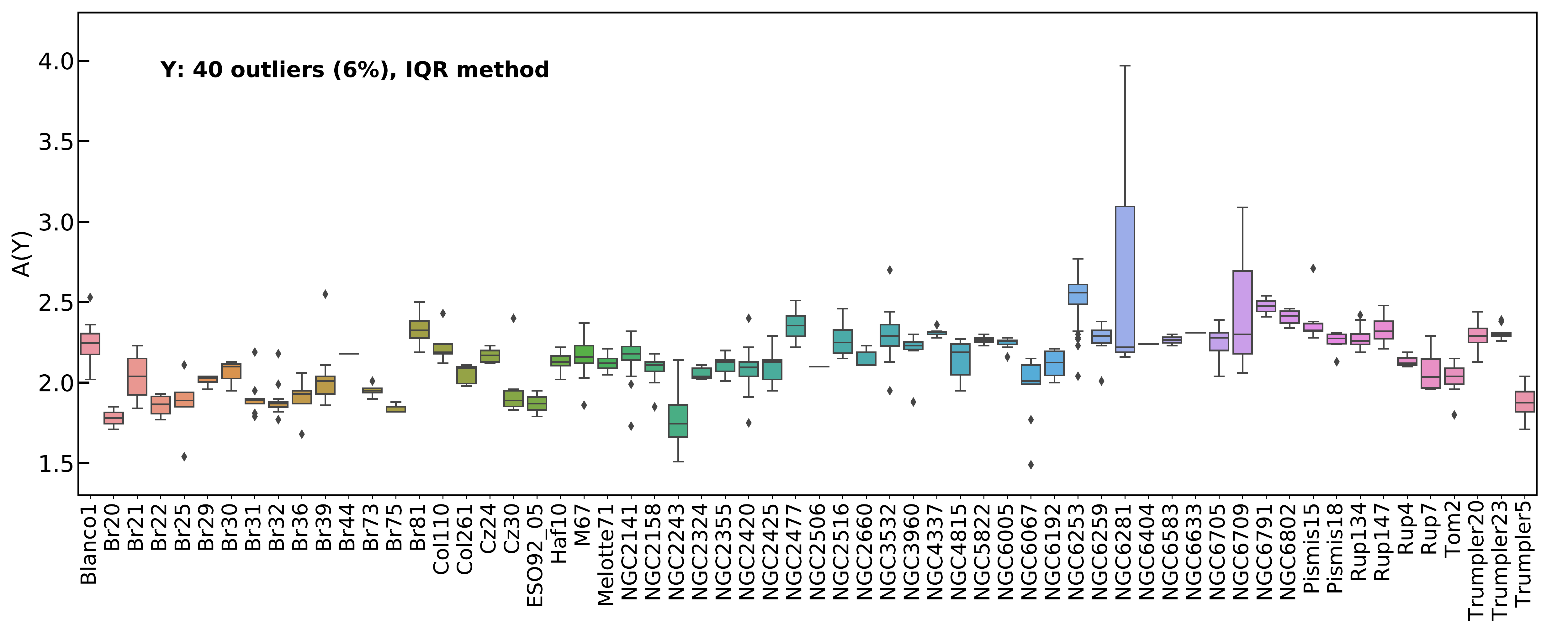}
  \caption{}
  \label{fig:outlier_y}
\end{subfigure}

\begin{subfigure}{\textwidth}
  \centering
  % include first image
  \includegraphics[scale=0.35]{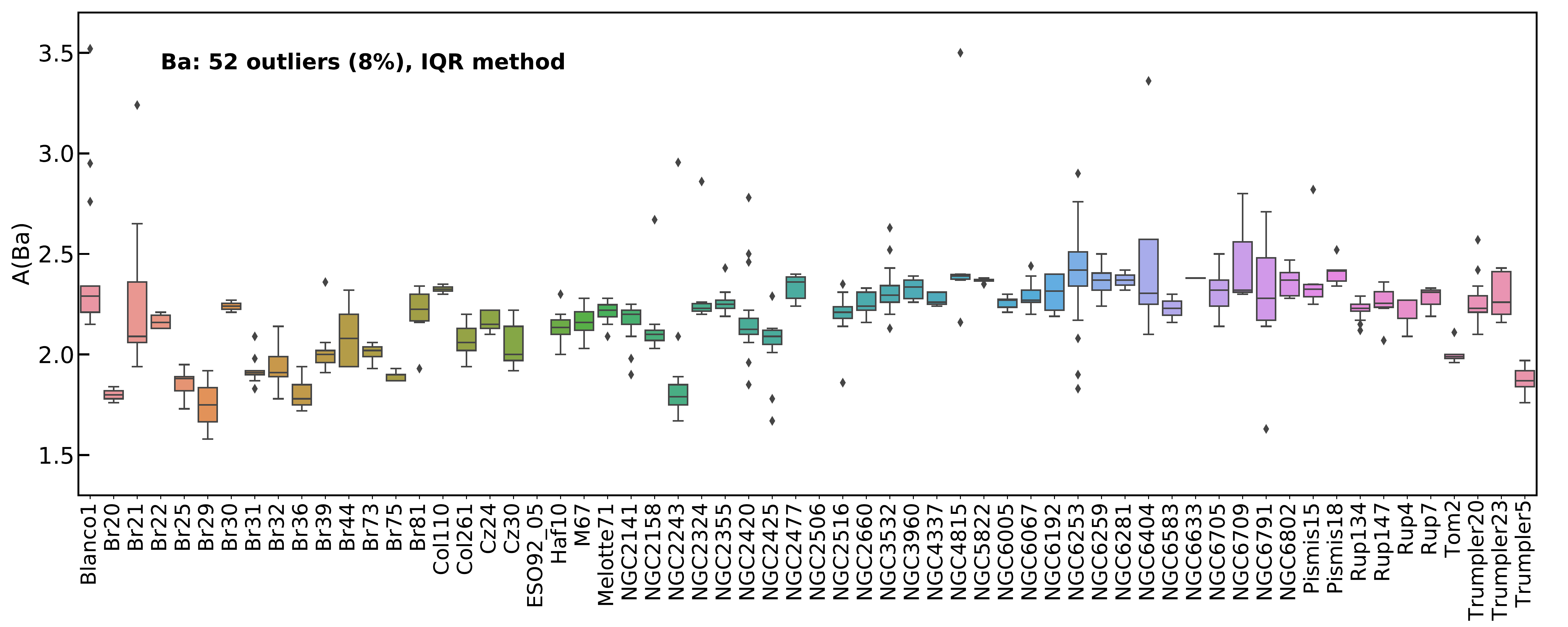}
  \caption{}
  \label{fig:outlier_ba}
\end{subfigure}

\caption{Box plots with the interquartile range of the abundance for each of the clusters with the outliers (observations that fall below Q1 - 1.5 IQR or above Q3 + 1.5 IQR). Figure \ref{fig:outlier_y} for Y2 and \ref{fig:outlier_ba} for Ba2.}
\label{fig:outliers}
\end{figure*}

%%%%%%%%%%%%%%%%%%%% tables %%%%%%%%%%%%%%%%%%%%%

\clearpage
\onecolumn

\begin{landscape}
\begin{longtable}{lrrrrrrrrrrrrrr}
\caption{\label{clusters_elh} Average [s-/H] and [s-/Fe] for our sample of open clusters}\\

\hline\hline
GES\_FLD & [Fe/H] &  Age (Gyr) & R$_{\rm GC}$ (kpc) & $[\ion{Y}{ii}$/H] & $[\ion{Zr}{i}$/H] & $[\ion{Ba}{ii}$/H] & $[\ion{La}{ii}$/H] & $[\ion{Ce}{ii}$/H] & $[\ion{Y}{ii}$/Fe] & $[\ion{Zr}{i}$/Fe] & $[\ion{Ba}{ii}$/Fe] & $[\ion{La}{ii}$/Fe] & $[\ion{Ce}{ii}$/Fe]\\
\hline
\endfirsthead
\caption{continued.}\\
\hline\hline
GES\_FLD & [Fe/H] &  Age (Gyr) & R$_{\rm GC}$ (kpc) & $[\ion{Y}{ii}$/H] & $[\ion{Zr}{i}$/H] & $[\ion{Ba}{ii}$/H] & $[\ion{La}{ii}$/H] & $[\ion{Ce}{ii}$/H] & $[\ion{Y}{ii}$/Fe] & $[\ion{Zr}{i}$/Fe] & $[\ion{Ba}{ii}$/Fe] & $[\ion{La}{ii}$/Fe] & $[\ion{Ce}{ii}$/Fe]\\
\hline
\endhead
\hline
%\endfoot
%%
  Blanco 1 & $-$0.03 & 0.1 & 8.3 & 0.0 & $-$0.04 & 0.05 & 0.12 & $-$0.06 & 0.02 & 0.02 & 0.09 & 0.13 & $-$0.02\\
  Berkeley 20 & $-$0.38 & 4.79 & 16.32 & $-$0.31 & $-$0.34 & $-$0.32 & $-$0.33 & $-$0.34 & 0.06 & 0.03 & 0.06 & 0.1 & 0.04\\
  Berkeley 21 & $-$0.21 & 2.14 & 14.73 & $-$0.05 & $-$0.21 & 0.06 &  & $-$0.04 & 0.16 & 0.01 & 0.29 &  & 0.19\\
  Berkeley 22 & $-$0.26 & 2.45 & 14.29 & $-$0.23 & $-$0.32 & 0.04 &  & $-$0.27 & 0.04 & $-$0.02 & 0.28 &  & 0.11\\
  Berkeley 25 & $-$0.25 & 2.45 & 13.81 & $-$0.2 & $-$0.25 & $-$0.27 & $-$0.27 & $-$0.3 & 0.11 & 0.05 & 0.07 & 0.09 & 0.03\\
  Berkeley 29 & $-$0.36 & 3.09 & 20.58 & $-$0.08 &  & $-$0.37 &  &  & 0.32 &  & $-$0.0 &  & \\
  Berkeley 30 & $-$0.13 & 0.3 & 13.25 & $-$0.03 & $-$0.08 & 0.12 & 0.13 & 0.06 & 0.12 & 0.09 & 0.27 & 0.28 & 0.21\\
  Berkeley 31 & $-$0.31 & 2.82 & 15.09 & $-$0.2 & $-$0.21 & $-$0.22 & $-$0.17 & $-$0.13 & 0.13 & 0.05 & 0.1 & 0.18 & 0.2\\
  Berkeley 32 & $-$0.29 & 4.9 & 11.14 & $-$0.23 & $-$0.18 & $-$0.18 & $-$0.15 & $-$0.21 & 0.06 & 0.1 & 0.1 & 0.13 & 0.07\\
  Berkeley 36 & $-$0.15 & 6.76 & 11.73 & $-$0.14 & $-$0.11 & $-$0.31 & 0.07 & $-$0.14 & 0.01 & 0.12 & $-$0.08 & 0.33 & 0.01\\
  Berkeley 39 & $-$0.14 & 5.62 & 11.49 & $-$0.12 & $-$0.07 & $-$0.14 & $-$0.05 & $-$0.07 & 0.02 & 0.07 & $-$0.0 & 0.09 & 0.06\\
  Berkeley 44 & 0.22 & 1.45 & 7.01 & 0.09 & 0.14 & $-$0.01 &  & 0.1 & $-$0.03 & $-$0.04 & $-$0.22 &  & $-$0.12\\
  Berkeley 73 & $-$0.26 & 1.41 & 13.76 & $-$0.16 & $-$0.19 & $-$0.11 & $-$0.12 & $-$0.09 & 0.11 & 0.08 & 0.14 & 0.12 & 0.17\\
  Berkeley 75 & $-$0.34 & 1.7 & 14.67 & $-$0.25 & $-$0.37 & $-$0.23 & $-$0.25 & $-$0.26 & 0.1 & 0.02 & 0.12 & 0.14 & 0.12\\
  Berkeley 81 & 0.22 & 1.15 & 5.88 & 0.24 & 0.22 & 0.13 & 0.3 & 0.3 & 0.02 & 0.01 & $-$0.08 & 0.09 & 0.07\\
  Collinder 110 & $-$0.1 & 1.82 & 10.29 & 0.09 & $-$0.02 & 0.2 & 0.11 & 0.14 & 0.17 & 0.07 & 0.31 & 0.17 & 0.23\\
  Collinder 261 & $-$0.05 & 6.31 & 7.26 & $-$0.04 & $-$0.03 & $-$0.05 & 0.04 & $-$0.08 & 0.01 & 0.02 & 0.0 & 0.07 & $-$0.0\\
  Czernik 24 & $-$0.11 & 2.69 & 12.29 & 0.08 & $-$0.01 & 0.04 & 0.16 & 0.07 & 0.19 & 0.09 & 0.15 & 0.26 & 0.18\\
  Czernik 30 & $-$0.31 & 2.88 & 13.78 & $-$0.21 & $-$0.2 & $-$0.07 &  & $-$0.08 & 0.11 & 0.14 & 0.26 &  & 0.24\\
  ESO92\_05 & $-$0.29 & 4.47 & 12.82 & $-$0.22 & $-$0.23 &  &  &  & 0.11 & 0.1 &  &  & \\
  Haffner 10 & $-$0.1 & 3.8 & 10.82 & 0.04 & $-$0.03 & 0.0 & 0.05 & 0.03 & 0.15 & 0.08 & 0.11 & 0.17 & 0.14\\
  M 67 & 0.0 & 4.27 & 8.96 & 0.0 & 0.0 & 0.0 & 0.02 & $-$0.01 & 0.03 & $-$0.01 & 0.02 & 0.03 & $-$0.01\\
  Melotte 71 & $-$0.15 & 0.98 & 9.87 & 0.04 & 0.03 & 0.11 & 0.05 & 0.05 & 0.19 & 0.14 & 0.22 & 0.21 & 0.17\\
  NGC 2141 & $-$0.04 & 1.86 & 13.34 & 0.11 & 0.03 & 0.07 & 0.14 & 0.07 & 0.15 & 0.07 & 0.12 & 0.16 & 0.11\\
  NGC 2158 & $-$0.15 & 1.55 & 12.62 & 0.02 & 0.0 & $-$0.02 & 0.04 & 0.01 & 0.18 & 0.17 & 0.15 & 0.19 & 0.18\\
  NGC 2243 & $-$0.45 & 4.37 & 10.58 & $-$0.33 & $-$0.39 & $-$0.35 & $-$0.38 & $-$0.38 & 0.12 & 0.09 & 0.11 & 0.08 & 0.09\\
  NGC 2324 & $-$0.18 & 0.54 & 12.08 & $-$0.03 & $-$0.1 & 0.11 & $-$0.11 & $-$0.03 & 0.15 & 0.08 & 0.29 & 0.07 & 0.16\\
  NGC 2355 & $-$0.13 & 1.0 & 10.11 & 0.02 & 0.04 & 0.13 & 0.06 & 0.06 & 0.12 & 0.14 & 0.23 & 0.16 & 0.16\\
  NGC 2420 & $-$0.15 & 1.74 & 10.68 & $-$0.05 & 0.01 & $-$0.01 & 0.04 & 0.0 & 0.13 & 0.16 & 0.16 & 0.19 & 0.16\\
  NGC 2425 & $-$0.12 & 2.4 & 10.92 & 0.0 & $-$0.05 & $-$0.03 & 0.09 & $-$0.01 & 0.11 & 0.06 & 0.1 & 0.19 & 0.12\\
  NGC 2477 & 0.14 & 1.12 & 8.85 & 0.22 & 0.18 & 0.19 & 0.13 & 0.17 & 0.08 & 0.08 & 0.04 & 0.02 & 0.06\\
  NGC 2506 & $-$0.34 & 1.66 & 10.62 & 0.01 &  &  &  &  & 0.35 &  &  &  & \\
  NGC 2516 & $-$0.04 & 0.24 & 8.32 & 0.05 &  & 0.02 & $-$0.31 &  & 0.1 &  & 0.06 &  & \\
  NGC 2660 & $-$0.05 & 0.93 & 8.98 & 0.06 & 0.01 & 0.13 & 0.12 & 0.14 & 0.11 & 0.06 & 0.18 & 0.17 & 0.19\\
  NGC 3532 & $-$0.03 & 0.4 & 8.19 & 0.09 & 0.09 & 0.11 & 0.07 & 0.2 & 0.1 & 0.09 & 0.12 & 0.09 & 0.19\\
  NGC 3960 & 0.0 & 0.87 & 7.68 & 0.15 & 0.09 & 0.22 & 0.18 & 0.16 & 0.15 & 0.09 & 0.22 & 0.17 & 0.16\\
  NGC 4337 & 0.24 & 1.45 & 7.45 & 0.21 & 0.19 & 0.15 & 0.15 & 0.18 & $-$0.03 & $-$0.06 & $-$0.1 & $-$0.11 & $-$0.08\\
  NGC 4815 & 0.08 & 0.37 & 7.07 & 0.05 & 0.08 & 0.27 & 0.02 & 0.18 & 0.06 & 0.1 & 0.25 & 0.06 & 0.17\\
  NGC 5822 & 0.02 & 0.91 & 7.69 & 0.18 & 0.1 & 0.25 & 0.12 & 0.14 & 0.16 & 0.08 & 0.24 & 0.11 & 0.13\\
  NGC 6005 & 0.22 & 1.26 & 6.51 & 0.16 & 0.21 & 0.14 & 0.06 & 0.16 & $-$0.05 & $-$0.01 & $-$0.08 & $-$0.16 & $-$0.06\\
  NGC 6067 & 0.03 & 0.13 & 6.78 & $-$0.02 & 0.09 & 0.16 & $-$0.02 & 0.02 & $-$0.02 & 0.16 & 0.23 & 0.01 & 0.05\\
  NGC 6192 & $-$0.08 & 0.24 & 6.73 & 0.02 & 0.02 & 0.18 & 0.12 & 0.09 & 0.11 & 0.1 & 0.27 & 0.22 & 0.18\\
  NGC 6253 & 0.34 & 3.24 & 6.88 & 0.36 & 0.21 & 0.25 & 0.31 & 0.25 & 0.02 & $-$0.03 & $-$0.09 & $-$0.01 & 0.02\\
  NGC 6259 & 0.18 & 0.27 & 6.18 & 0.2 & 0.16 & 0.25 & 0.17 & 0.28 & 0.03 & $-$0.03 & 0.06 & $-$0.01 & 0.1\\
  NGC 6281 & $-$0.04 & 0.51 & 7.81 & 0.1 & 0.04 & 0.25 & 0.14 & 0.14 & 0.14 & 0.09 & 0.3 & 0.15 & 0.18\\
  NGC 6404 & 0.01 & 0.1 & 5.85 & 0.15 & $-$0.14 & 0.12 & 0.08 & 0.13 & 0.2 & $-$0.11 & 0.17 & 0.06 & 0.12\\
  NGC 6583 & 0.22 & 1.2 & 6.32 & 0.18 &  & 0.11 &  & 0.23 & $-$0.04 &  & $-$0.11 &  & 0.02\\
  NGC 6633 & $-$0.03 & 0.69 & 8.0 & 0.1 &  & 0.19 & 0.03 &  & 0.09 &  & 0.18 & 0.02 & \\
  NGC 6705 & 0.03 & 0.31 & 6.46 & 0.15 & 0.08 & 0.19 & 0.14 & 0.18 & 0.11 & 0.05 & 0.15 & 0.11 & 0.15\\
  NGC 6709 & $-$0.02 & 0.19 & 7.6 & 0.03 & 0.01 & 0.16 & 0.1 & 0.07 & 0.08 & 0.08 & 0.2 & 0.17 & 0.14\\
  NGC 6791 & 0.22 & 6.31 & 7.94 & 0.38 &  & 0.17 & 0.21 & 0.34 & 0.06 &  & $-$0.02 & $-$0.11 & 0.12\\
  NGC 6802 & 0.14 & 0.66 & 7.14 & 0.32 & 0.2 & 0.24 & 0.13 & 0.22 & 0.19 & 0.06 & 0.11 & $-$0.0 & 0.08\\
  Pismis 15 & 0.02 & 0.87 & 8.62 & 0.24 & 0.1 & 0.19 & 0.24 & 0.18 & 0.2 & 0.08 & 0.15 & 0.19 & 0.14\\
  Pismis 18 & 0.14 & 0.58 & 6.94 & 0.19 & 0.17 & 0.27 & 0.22 & 0.19 & 0.04 & 0.02 & 0.14 & 0.09 & 0.04\\
  Ruprecht 134 & 0.27 & 1.66 & 6.09 & 0.17 & 0.21 & 0.11 & 0.12 & 0.2 & $-$0.09 & $-$0.06 & $-$0.16 & $-$0.15 & $-$0.06\\
  Ruprecht 147 & 0.12 & 3.02 & 8.05 & 0.12 &  & 0.1 & 0.06 &  & 0.0 &  & $-$0.03 & $-$0.06 & \\
  Ruprecht 4 & $-$0.13 & 0.85 & 11.68 & 0.05 & $-$0.09 & 0.09 & 0.11 & 0.11 & 0.18 & 0.06 & 0.22 & 0.25 & 0.24\\
  Ruprecht 7 & $-$0.24 & 0.23 & 13.11 & $-$0.01 & $-$0.07 & 0.16 & $-$0.04 & $-$0.04 & 0.23 & 0.19 & 0.4 & 0.2 & 0.18\\
  Tombaugh 2 & $-$0.24 & 1.62 & 15.76 & $-$0.03 & $-$0.06 & $-$0.14 & 0.0 & $-$0.27 & 0.2 & 0.09 & 0.14 & 0.24 & 0.03\\
  Trumpler 20 & 0.13 & 1.86 & 7.18 & 0.21 & 0.17 & 0.12 & 0.18 & 0.15 & 0.07 & 0.06 & $-$0.0 & 0.04 & 0.02\\
  Trumpler 23 & 0.2 & 0.71 & 6.27 & 0.2 & 0.19 & 0.17 & 0.17 & 0.19 & $-$0.0 & $-$0.01 & $-$0.02 & $-$0.04 & $-$0.02\\
  Trumpler 5 & $-$0.35 & 4.27 & 11.21 & $-$0.21 & $-$0.3 & $-$0.24 & $-$0.19 & $-$0.24 & 0.14 & 0.05 & 0.11 & 0.17 & 0.11\\
\hline
\footnote{[Fe/H] are from \citet{Randich_2022}, except clusters not present in that work for which it was calculated in this work.  R$_{\rm GC}$ and age from \citet{CG20}.} \\

\end{longtable}
\end{landscape}

\begin{table*}

\caption{WLS fitting coefficients of the relation [s-/Fe] = $m_{1}\cdot$Age + c for the Open Clusters in the three regions.}              % title of Table
\label{age_elfe1_coeffs_table}      % is used to refer this table in the text
\centering  
% used for centering table
\scalebox{0.9}{
\begin{tabular}{lrcccccccr}          % centered columns (4 columns)
\hline\hline % inserts double horizontal lines
 & & R$_{\rm GC}$ > 9 & & & 7 $\leq$ R$_{\rm GC}$ $\leq$ 9  & & & R$_{\rm GC}$ < 7  &\\
\hline
 & $m_{1}$ & c & PCC & $m_{1}$ & c & PCC & $m_{1}$ & c & PCC \\    % table heading
\hline % inserts single horizontal line
  $[\ion{Y}{ii}/\ion{Fe}]$ & $-$0.016$\pm$0.007 & 0.169$\pm$0.021 & $-$0.38 & $-$0.034$\pm$0.012 & 0.153$\pm$0.019 & $-$0.57& $-$0.016$\pm$0.017 & 0.025$\pm$0.027 & $-$0.29\\
  
  $[\ion{Zr}{i}/\ion{Fe}]$ & $-$0.014$\pm$0.007 & 0.126$\pm$0.019 & $-$0.40 & $-$0.019$\pm$0.009 & 0.090$\pm$0.015 & $-$0.52& $-$0.036$\pm$0.018 & 0.055$\pm$0.027 & $-$0.55\\
  
  $[\ion{Ba}{ii}/\ion{Fe}]$ & $-$0.043$\pm$0.006 & 0.266$\pm$0.019 & $-$0.83 & $-$0.033$\pm$0.014 & 0.141$\pm$0.033 & $-$0.48& $-$0.104$\pm$0.031 & 0.130$\pm$0.048 & $-$0.73\\
  
  $[\ion{La}{ii}/\ion{Fe}]$ & $-$0.021$\pm$0.006 & 0.228$\pm$0.017 & $-$0.65 & $-$0.032$\pm$0.013 & 0.162$\pm$0.026 & $-$0.57& $-$0.061$\pm$0.027 & 0.061$\pm$0.024 & $-$0.61\\
  
  $[\ion{Ce}{ii}/\ion{Fe}]$ & $-$0.020$\pm$0.003 & 0.191$\pm$0.007 & $-$0.80 & $-$0.036$\pm$0.012 & 0.145$\pm$0.024 & $-$0.62& $-$0.057$\pm$0.027 & 0.110$\pm$0.032 & $-$0.57\\
\hline
\end{tabular}}
\label{tab_coefficients}
\end{table*}

\begin{table*}
\caption{Weighted multilinear regressions of 3 variables fitting coefficients of the relation [s-/$\alpha$] = $m_{1}\cdot$Age + $m_{2} \cdot$ R$_{\rm GC}$ + $m_{3} \cdot$ [Fe/H] + c for the Open Clusters in every region and the coefficients of the inverted stellar dating relation Age = $m_{1}'\cdot$ [s-/$\alpha$] + $m_{2}'\cdot$ R$_{\rm GC}$ + $m_{3}'\cdot$ [Fe/H] + c'}              % title of Table
\label{tab:age_s_alpha_MLR_3_vbles}      % is used to refer this table in the text
\centering      % used for centering table
\scalebox{0.9}{
\begin{tabular}{lrrrrrrrrr}      
\hline\hline % inserts double horizontal lines
       & & & & R$_{\rm GC}$ > 9 kpc& & & & & \\
\hline % inserts single horizontal line
[s-/$\alpha$] & $m_{1}$ $\pm \Delta m_{1}$ & $m_{2}$ $\pm \Delta m_{2}$ & $m_{3}$ $\pm \Delta m_{3}$ & c $\pm \Delta c$  & $R_{adj}^2$ & c' & $m_{1}$' & $m_{2}$' & $m_{3}$' \\    % table heading
\hline
  $[$Y/Mg$]$ & $-$0.036$\pm$ 0.007 & $-$0.005$\pm$0.006  & 0.179$\pm$0.103 & 0.319$\pm$0.076 & 0.61 & 8.861 & $-$27.778 & $-$0.139 & 4.972\\
  $[$Y/Al$]$ & $-$0.033$\pm$0.007 & $-$0.005$\pm$0.006 & 0.320$\pm$0.115 & 0.338$\pm$0.076 & 0.59 & 10.242 & $-$30.303 & $-$0.152 & 9.697\\
  $[$Y/Si$]$ & $-$0.025$\pm$0.009 & 0.004$\pm$0.006 & 0.176$\pm$0.121 & 0.180$\pm$0.081 & 0.37 & 7.2 & $-$40.0 & 0.16 & 7.04\\
  $[$Y/Ca$]$ & $-$0.015$\pm$0.008 & 0.016$\pm$0.005 & 0.352$\pm$0.129 & $-$0.003$\pm$0.068 & 0.51 & $-$0.2 & $-$66.667 & 1.067 & 23.467\\
  $[$Y/Ti$]$ & $-$0.042$\pm$0.007  & $-$0.001$\pm$0.007  & 0.142$\pm$0.119 & 0.242$\pm$0.084 & 0.64 & 5.762 & $-$23.81 & $-$0.024 & 3.381\\

  \hline
  $[$Ba/Mg$]$ & $-$0.102$\pm$0.005  & 0.003$\pm$0.007 & 0.046$\pm$0.010 & 0.425$\pm$0.092 & 0.95 & 4.167 & $-$9.804 & 0.029 & 0.451\\
  $[$Ba/Al$]$ & $-$0.074$\pm$0.009 & $-$0.013$\pm$0.007 & $-$0.063$\pm$0.121 & 0.490$\pm$0.094 & 0.75 & 6.622 & $-$13.514 & $-$0.176 & $-$0.851\\
  $[$Ba/Si$]$ & $-$0.054$\pm$0.010 & $-$0.020$\pm$0.007 & $-$0.145$\pm$0.134 & 0.496$\pm$0.091 & 0.55 & 9.185 & $-$18.519 & $-$0.37 & $-$2.685\\
  $[$Ba/Ca$]$ & $-$0.055$\pm$0.006 & 0.021$\pm$0.000 & 0.002$\pm$0.000 & $-$0.001$\pm$0.000 & 0.82 & $-$0.018 & $-$18.182 & 0.382 & 0.036\\
  $[$Ba/Ti$]$ & $-$0.083$\pm$0.011 & $-$0.028$\pm$0.008 & $-$0.289$\pm$0.186 & 0.623$\pm$0.108 & 0.68 & 7.506 & $-$12.048 & $-$0.337 & $-$3.482\\
\hline
        & & & & 7 < R$_{\rm GC}$ < 9 kpc & & & & & \\
\hline
  $[$Y/Mg$]$ & $-$0.044$\pm$0.015 & $-$0.019$\pm$0.034 & 0.256$\pm$0.314 & 0.379$\pm$0.272 & 0.32 & 8.614 & $-$22.727 & $-$0.432 & 5.818\\
  $[$Y/Al$]$ & $-$0.051$\pm$0.014 & 0.025$\pm$0.001 & 0.002$\pm$0.000 & 0.004$\pm$0.000 & 0.42 & 0.078 & $-$19.608 & 0.49 & 0.039\\
  $[$Y/Si$]$ & $-$0.023$\pm$0.010 & 0.015$\pm$0.003 & $-$0.001$\pm$0.000 & 0.002$\pm$0.000 & 0.18 & 0.087 & $-$43.478 & 0.652 & $-$0.043\\
  $[$Y/Ca$]$ & $-$0.022$\pm$0.011 & $-$0.034$\pm$0.026 & $-$0.214$\pm$0.163 & 0.402$\pm$0.213 & 0.18 & 18.273 & $-$45.455 & $-$1.545 & $-$9.727\\
  $[$Y/TI$]$ & $-$0.022$\pm$0.016 & $-$0.001$\pm$0.038 & 0.191$\pm$0.295 & 0.162$\pm$0.304 & $-$0.03 & 7.364 & $-$45.455 & $-$0.045 & 8.682\\
 
\hline
  $[$Ba/Mg$]$ & $-$0.035$\pm$0.015 & $-$0.050$\pm$0.043 & $-$0.436$\pm$0.414 & 0.594$\pm$0.343 & 0.38 & 16.971 & $-$28.571 & $-$1.429 & $-$12.457\\
  $[$Ba/Al$]$ & $-$0.053$\pm$0.028 & $-$0.157$\pm$0.074 & $-$1.429$\pm$0.530 & 1.596$\pm$0.573 & 0.57 & 30.113 & $-$18.868 & $-$2.962 & $-$26.962\\
  $[$Ba/Si$]$ & $-$0.036$\pm$0.020 & $-$0.104$\pm$0.053 & $-$1.039$\pm$0.419 & 1.060$\pm$0.410 & 0.56 & 29.444 & $-$27.778 & $-$2.889 & $-$28.861\\
  $[$Ba/Ca$]$ & $-$0.028$\pm$0.013 & $-$0.070$\pm$0.032 & $-$0.479$\pm$0.273 & 0.716$\pm$0.258 & 0.46 & 25.571 & $-$35.714 & $-$2.5 & $-$17.107\\
  $[$Ba/Ti$]$ & $-$0.032$\pm$0.017 & $-$0.094$\pm$0.045 & $-$0.863$\pm$0.329 & 0.983$\pm$ 0.353 & 0.45 & 30.719 & $-$31.25 & $-$2.938 & $-$26.969\\
  
\hline
        & & & & R$_{\rm GC}$ < 7 kpc & & & & & \\
\hline
  $[$Y/Mg$]$ & $-$0.015$\pm$0.074 & 0.191$\pm$0.115 & 0.641$\pm$0.786 & $-$1.365$\pm$0.789 & 0.25 & $-$91.0 & $-$66.667 & 12.733 & 42.733\\
  $[$Y/Al$]$ & $-$0.103$\pm$0.066 & 0.050$\pm$0.100 & 0.780$\pm$0.726 & $-$0.307$\pm$0.683 & 0.12 & $-$2.981 & $-$9.709 & 0.485 & 7.573\\
  $[$Y/Si$]$ & $-$0.015$\pm$0.056 & 0.133$\pm$0.082 & 0.011$\pm$0.620 & $-$0.865$\pm$0.567 & 0.12 & $-$57.667 & $-$66.667 & 8.867 & 0.733\\
  $[$Y/Ca$]$ & $-$0.024$\pm$0.037 & 0.108$\pm$0.067 & 0.194$\pm$0.355 & $-$0.637$\pm$0.450 & $-$0.01 & $-$26.542 & $-$41.667 & 4.5 & 8.083\\
  $[$Y/Ti$]$ & $-$0.070$\pm$0.064 & 0.114$\pm$0.107 & 0.297$\pm$0.654 & $-$0.609$\pm$0.711 & $-$0.04 & $-$8.7 & $-$14.286 & 1.629 & 4.243\\

\hline

  $[$Ba/Mg$]$ & $-$0.031$\pm$0.033 & 0.254$\pm$0.041 & $-$0.173$\pm$0.279 & $-$1.606$\pm$0.253 & 0.76 & $-$51.806 & $-$32.258 & 8.161 & $-$5.581\\
  $[$Ba/Al$]$ & $-$0.151$\pm$0.035 & 0.169$\pm$0.043 & 0.507$\pm$0.367  & $-$0.976$\pm$0.299 & 0.84 & $-$6.464 & $-$6.623 & 1.119 & 3.358\\
  $[$Ba/Si$]$ & $-$0.053$\pm$0.029 & 0.235$\pm$0.042 & $-$0.403$\pm$0.287 & $-$1.406$\pm$0.277 & 0.89 & $-$26.528 & $-$18.868 & 4.434 & $-$7.604\\
  $[$Ba/Ca$]$ & $-$0.009$\pm$0.043 & 0.111$\pm$0.056 & $-$0.686$\pm$0.307 & $-$0.517$\pm$0.354 & 0.76 & $-$57.444 & $-$111.111 & 12.333 & $-$76.222\\
  $[$Ba/Ti$]$ & $-$0.137$\pm$0.029 & 0.237$\pm$0.049 & 0.239$\pm$0.307 & $-$1.333$\pm$0.330 & 0.88 & $-$9.73 & $-$7.299 & 1.73 & 1.745\\  
 
\hline
        & & & & global &  & & & & \\
\hline
  $[$Y/Mg$]$ & $-$0.036$\pm$0.011 & 0.006$\pm$0.010 & $-$0.166$\pm$0.122 & 0.129$\pm$0.079 & 0.16 & 3.583 & $-$27.778 & 0.167 & $-$4.611\\
  $[$Y/Al$]$ & $-$0.039$\pm$0.007 & $-$0.001$\pm$0.005 & $-$0.190$\pm$0.060 & 0.191$\pm$0.032 & 0.54 & 4.897 & $-$25.641 & $-$0.026 & $-$4.872\\
  $[$Y/Si$]$ & $-$0.026$\pm$0.006 & 0.010$\pm$0.005 & $-$0.125$\pm$0.062 & 0.035$\pm$0.032 & 0.34 & 1.346 & $-$38.462 & 0.385 & $-$4.808\\
  $[$Y/Ca$]$ & $-$0.019$\pm$0.006 & 0.012$\pm$0.004 & 0.075$\pm$0.068 & 0.003$\pm$0.038  & 0.29 & 0.158 & $-$52.632 & 0.632 & 3.947\\
  $[$Y/Ti$]$ & $-$0.035$\pm$0.007 & 0.006$\pm$0.007 & 0.054$\pm$0.086 & 0.114$\pm$0.060 & 0.40 & 3.257 & $-$28.571 & 0.171 & 1.543\\
  
\hline
  $[$Ba/Mg$]$ & $-$0.103$\pm$0.006 & 0.038$\pm$0.003 & 0.025$\pm$0.021 & 0.050$\pm$0.046 & 0.86 & 0.485 & $-$9.709 & 0.369 & 0.243\\
  $[$Ba/Al$]$ & $-$0.089$\pm$0.012 & $-$0.014$\pm$0.009 & $-$0.603$\pm$0.134 & 0.448$\pm$0.083 & 0.55 & 5.034 & $-$11.236 & $-$0.157 & $-$6.775\\
  $[$Ba/Si$]$ & $-$0.061$\pm$0.009  & $-$0.013$\pm$0.007 & $-$0.536$\pm$0.112 & 0.340$\pm$0.068 & 0.51 & 5.574 & $-$16.393 & $-$0.213 & $-$8.787\\
  $[$Ba/Ca$]$ & $-$0.055$\pm$0.005 & 0.020$\pm$0.002 & 0.009$\pm$0.014 & 0.015$\pm$0.030 & 0.84 & 0.273 & $-$18.182 & 0.364 & 0.164\\
  $[$Ba/Ti$]$ & $-$0.066$\pm$0.008 & $-$0.017$\pm$0.008 & $-$0.477$\pm$0.102 & 0.397$\pm$0.076 & 0.56 & 6.015 & $-$15.152 & $-$0.258 & $-$7.227\\
\hline
\end{tabular}}
\end{table*}

\end{appendix}

\end{document}